\documentclass[aps,prd,reprint,superscriptaddress, floatfix]{revtex4-2}

\usepackage{color}
\usepackage{amssymb}
\usepackage{graphicx,color}
\usepackage[usenames,dvipsnames]{xcolor}
\usepackage[section]{placeins}
\usepackage{amsmath}
\usepackage{subfigure}
\usepackage[normalem]{ulem}
\usepackage{booktabs}
\usepackage{xcolor}
\usepackage{epsfig,graphics,color,graphicx,amsmath}

\colorlet{darkgreen}{green!50!black}
\colorlet{brightyellow}{yellow!75!red}
\colorlet{orange}{red!50!yellow}
\colorlet{darkblue}{blue!60!black}
\colorlet{darkred}{red!80!black}

\usepackage{soul}

\usepackage{mathtools}
\usepackage{mathrsfs}
\usepackage{bm}
\usepackage{relsize}
\usepackage{indentfirst}

\usepackage{xcolor}

\newcommand{\cd}{\makebox[0.08cm]{$\cdot$}}

\begin{document}

\title{Three-boson bound states in Minkowski space with contact interactions}
\author{E.~Ydrefors}
\affiliation{Instituto Tecnol\'ogico de Aeron\'autica,  DCTA, 
12228-900 S\~ao Jos\'e dos Campos,~Brazil}
\affiliation{Universit\'{e} Paris-Saclay, CNRS/IN2P3, IJCLab, 91405 Orsay, France}
\author{J.H.~Alvarenga Nogueira}
\affiliation{Instituto Tecnol\'ogico de Aeron\'autica,  DCTA, 
12228-900 S\~ao Jos\'e dos Campos,~Brazil}
\affiliation{Dipartimento di Fisica, Universit\`a di Roma ``La Sapienza" \\
INFN, Sezione di Roma ``La Sapienza"
Piazzale A. Moro 5 - 00187 Roma, Italy}
\author{V.A.~Karmanov}
\affiliation{Lebedev Physical Institute, Leninsky Prospekt 53, 119991 Moscow, Russia}
\author{T.~Frederico}
\affiliation{Instituto Tecnol\'ogico de Aeron\'autica,  DCTA, 
12228-900 S\~ao Jos\'e dos Campos,~Brazil}
\date{\today}

\begin{abstract}
The structure of the three-boson bound state in Minkowski space is studied for a model with contact interaction. 
  The Faddeev-Bethe-Salpeter equation is solved both in Minkowski and Euclidean spaces. The results are in fair agreement for comparable quantities,
  like the  transverse amplitude obtained when  the longitudinal constituent momenta of the light-front valence wave function are integrated out.
  The Minkowski space solution is obtained numerically by using a recently proposed method  based on the  direct integration   over the singularities of the propagators and interaction kernel of the four-dimensional integral equation.  
  The complex singular structure of the Faddeev components of the Bethe-Salpeter vertex function  for space and time-like momenta 
  in an example of a  Borromean system is investigated in detail.
Furthermore, the  transverse amplitude  is studied as a mean to access  the  double-parton transverse momentum distribution. Following that, 
we show that  the two-body short-range correlation contained in the valence wave function is evidenced when the pair has a large relative momentum in 
a back-to-back configuration, where  one of the Faddeev components of the  Bethe-Salpeter 
amplitude dominates over the others. In this situation a  power-law behavior is derived and confirmed numerically.
    
\end{abstract}
\maketitle


\section{Introduction}\label{Sec:intr}

The Bethe-Salpeter (BS) approach is an important and efficient tool to investigate relativistic few-body systems. 
Solving the BS equation  with a realistic interaction, especially for a three-body system, is technically a rather complicated problem. However, the principal qualitative properties can be understood through models that retain the main features of the physical system. One of these models, fundamental in nuclear physics and described (in the two-body case) in any textbook, is the zero-range interaction.

The three-body BS equation in Minkowski space with zero-range interaction was derived in Ref.~\cite{tobias1} in 1992.  Later on, in 2017, the equation  was solved in  Euclidean space \cite{ey3b} for the first time and 
then, quite recently, directly in  Minkowski space \cite{Ydrefors19_3b_Mink}. Concerning the Euclidean space solution,  the reasons of this time lag was due to the fact that, though the BS equation was given in  Ref.~\cite{tobias1} in a simple and transparent form, as it was there presented the equation did not 
allow to make the Wick rotation directly. Whether the rotating integration contour crosses the singularities or not,  this depends on the point around which it is rotated. To determine the safe point one must make a shift of variables in the BS equation~\cite{tobias1}, as proposed in Ref.~\cite{ey3b}. This was the key to success.  As for the Minkowski space solution~\cite{Ydrefors19_3b_Mink}, the methods were absent until recently. 
In Ref.~\cite{Ydrefors19_3b_Mink} the method developed in  Ref.~\cite{ck2b} was used.

The aforementioned  method is based on the direct integration of  the singularities of the propagators and interaction kernel~\cite{ck2b}. 
It does not resort to the Nakanishi integral representation \cite{Nakanishi63, Nakanishi69} and light-front (LF) projection. 
In the present paper we will follow this method and explore it for obtaining information on the structure of relativistic three-body systems. 

However, the corresponding equation in the light-front dynamics (LFD) was derived and solved already in Ref.~\cite{tobias1}. Then the stability of the solution was thoroughly explored in Ref.~\cite{ck3b}.  Finding the solution of the BS equation fully in Minkowski space is rather important for applications when used to calculate  observables like parton distributions and  electromagnetic form factors (see e.g.~Ref.~\cite{ck-trento09}). The Euclidean solution, though it provides the bound state  spectrum, requires a  careful analytical extension of the Euclidean BS amplitude, and for large enough momentum transfers overlapping cuts turns the task cumbersome, preventing to explore the whole range of momentum transfers. Besides that, the comparison between the BS and the LFD solutions  provides valuable information about the structure of the system, i.e., regarding contribution of the higher Fock components etc.

 As it was shown in Ref.~\cite{ey3b}, the effect coming from higher Fock components on the binding energy and transverse amplitude is huge, even for weakly bound states. This is different from the two-body case, where the truncation at the valence state does not present such a dramatic effect (see e.g.~Refs.~\cite{FrePRD14,Ji-Tokunaga12}).  This difference in a three-body system is explained by  the contribution of  effective three-body forces of relativistic origin, as investigated in Ref.~\cite{Karmanov:2008bx}.  Noteworthy that the BS equation for three bosons has a kernel analogous to the contribution provided by the quark exchange diagrams in quark-diquark models in the constituent quark picture~\cite{Eichmannreview}, making even more appealing the outcomes of the Minkowski space approach to be presented as follows.      
                                   
The conclusion that the effect coming from higher Fock components is sizable leads to 
raise doubts regarding the range of validity of valence inspired models, which are widely applied to hadron physics, as they might be inappropriate  
to describe certain features of the bound state dynamics, particularly for three-body systems. It is worth 
mentioning that even for two-boson bound states the contributions coming from higher Fock components, as shown by the calculations \cite{FrePRD14,Hwang04}, can constitute more than 30\% of the normalization.    
The BS equation and LFD  approaches  have already been used as a suitable framework  
in phenomenological applications. For instance,   the calculation of the LF amplitudes in a simplified pion model with strongly bound constituent quarks was done
through the solution of the BS equation directly 
in Minkowski space \cite{dePaulaEPJC} and also the final state interaction in heavy meson decays
 was studied using a relativistic LF model~\cite{FBS2017,JHEP14}.

As  mentioned,  the comparison of the binding energies calculated within 
LFD and BS equation for a one-boson exchange kernel presents a significant
discordance~\cite{Karmanov:2008bx}, unlike what happens for 
two-body systems \cite{mangin}. 
In Ref.~\cite{Karmanov:2008bx} there was found an increasing effect of the three-body forces as the exchanged boson mass $\mu$ grows, what is relevant for the zero-range case, which corresponds effectively to $\mu \to \infty$.
Although that work was quite instructive, the three-body forces were taken into account only perturbatively, producing a significant contribution to the bound state energy, what indicates the necessity to go beyond perturbation theory. It is essential to obtain the non-perturbative 
solution of the three-body BS and LFD equations, including three-body forces, 
in order to have a thorough understanding of the physical system.

In the non-relativistic approach, within the Schr\"odinger equation, it is well known that the binding energy of a  three-boson system with the two-body zero-range interaction is not bound from below, what is known as the Thomas collapse~\cite{thomas}.
As shown in Ref.~\cite{tobias1}, and further explored numerically in Ref.~\cite{ck3b}, the relativistic effects result in an effective repulsion at small distances that prevents the Thomas collapse in the relativistic case. This result was found for the 
valence truncation, within the LFD framework.  Therefore, exploring the complete amplitude by 
means of the BS equation, which includes higher Fock contributions, is  necessary to describe such a relativistic three-body system. 

Furthermore, the approach for three-boson systems allows one to explore within the relativistic context a wide and important field of research that is already very well established non-relativistically, known as the Efimov physics~\cite{Tobiasreview,Efimov1970}. The three-body approach developed here paves the way to explore many interesting relativistic phenomena  and it is expected to bring more remarkable outcomes as further  studies are done.

This paper is devoted to a detailed study of the Minkowski space solution of the three-boson Faddeev-BS equation in the case of the two-body zero-range interaction. 
 As the goal  is to address the zero-range interaction case,
a major point is the influence of relativistic effects 
on the stability of the three-body system and the impact on its structure.
 To accomplish such a goal we focus on Borromean systems and the Faddeev-BS equation is solved both in Minkowski and Euclidean spaces. 
The choice made for Borromean states  simplifies the computations in Minkowski space, as the bound state pole is absent in the two-body scattering amplitude, which is an 
 input to the kernel of the Faddeev-BS equation.     
 
 The Minkowski space solution is obtained by the direct integration of  the singularities of the propagators and interaction kernel~\cite{Ydrefors19_3b_Mink}, what allows
  to explore in the space and time-like momenta regions the complex singular structure of the Faddeev components of the BS vertex function of such a Borromean state.   
 We study in detail the numerical solutions by showing that both methods produce results 
 in fair agreement for the Faddeev component of the transverse amplitude obtained from the corresponding component of the valence wave function, after integration over the longitudinal LF momentum fractions.
%
 In addition, the double-parton content of the transverse amplitude  is studied, and
we evidenced the two-body short-range correlation  contained in the valence wave function. The kinematical condition to expose the pair  short-range correlation
was set for large relative momentum in a back-to-back configuration, in such situation the Faddeev component of the  BS amplitude that brings the  pair interaction is the dominant one. We also found, as expected for large relative momentum, a power-law behavior, that was  confirmed numerically.

The paper is organized as follows. The theoretical formalism for the two-body scattering amplitude, 3-body BS equation and  transverse amplitudes is outlined in Secs.~\ref{Sec:two_body_amp}-\ref{Sec:TA}. In Sec.~\ref{Sec:wick_rotation} the Wick rotation of the three-body BS amplitude is revisited to clarify that the transverse amplitude is
independent of that. In Sec.~\ref{Sec:Res} the numerical results are presented and discussed. The conclusions are then drawn in Sec.~\ref{Sec:Concl}. Some of the more lengthy derivations, and also a brief summary of the numerical methods, are available in appendices.  

\section{Two-body scattering amplitude}\label{Sec:two_body_amp}

\begin{figure}[!bth]
\centering
\includegraphics[scale=0.4]{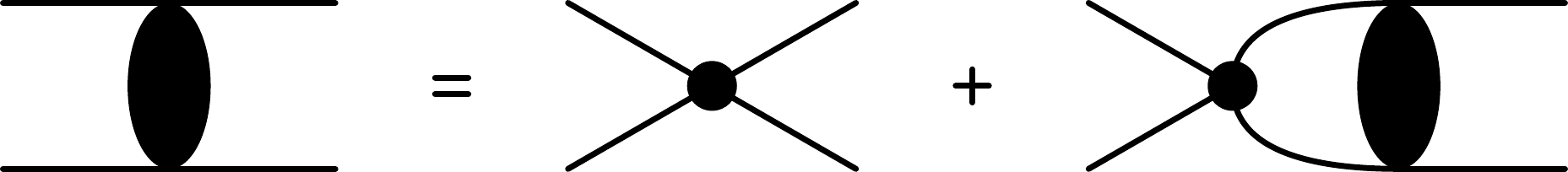} 
\caption{Diagrammatic representation of the integral equation for the two-body scattering amplitude \eqref{eq37}\label{Fig:two-body-amp} of Ref.~\cite{tobias1}.}
\end{figure}

For the contact interaction (with the four-leg vertex $i\lambda$), the two-body amplitude ${\cal F}(M^2_{12})$  is determined by the equation  shown graphically in Fig.~\ref{Fig:two-body-amp} (see also Ref.~\cite{tobias1}). 
Iterating, we find that the first contribution is simply $i \lambda$, the second one is $(i \lambda)^2{\cal B}$,  where ${\cal B}$ is the amputated from $(i \lambda)^2$ the bubble graph,  etc.
That is:
\begin{equation}
\begin{aligned}
\label{eq37}
i{\cal F}(M^2_{12})&= i \lambda+(i \lambda)^2{\cal B}+(i \lambda)^3{\cal B}^2+\ldots=\\
& \frac{i \lambda}{1-(i \lambda){\cal B}(M^2_{12})}= \frac{1}{(i \lambda)^{-1}-{\cal B}(M^2_{12})},
\end{aligned}
\end{equation}
or
\begin{equation}\label{eq37a}
{\cal F}(M^2_{12})= \frac{1}{i[(i \lambda)^{-1}-{\cal B}(M^2_{12})]},
\end{equation}
where 
\begin{equation}\label{eq38}
\begin{aligned}
{\cal B}&(M^2_{12})=\\
&\int\frac {d^4k}{(2\pi)^4}\frac{i}{(k^2-m^2+i\epsilon)}\frac{i}{[(k-P)^2-m^2+i\epsilon]}. 
\end{aligned}
\end{equation}
Here $m$ denotes the boson mass, and $P$ is the total four-momentum of the two-body system, 
$P^2\equiv M^2_{12}$, where $M^2_{12}$ denotes the squared effective off-shell mass.

The loop integration (\ref{eq38}) has a log-type ultraviolet divergence that has to be regularized and renormalized by fixing the scattering amplitude
at some physical value, which  will be given by the bound state pole or scattering length. Although it is a well known and standard procedure, we will present it in the following for the sake of completeness as well as to fix our notation.

\subsection{Normalization of the scattering amplitude}
 
 In the derivation of the two-body amplitude, we follow the definitions and normalization of Ref.~\cite{IZ}. According to it, the partial wave amplitude of angular momentum $L$ is defined as 
\begin{equation}\label{fpw2}
 F_L(k_v)=\frac{1}{32\pi}\int_{-1}^1dz \; P_L(z) F(k_v, z)\, ,
\end{equation}
where $k_v = |\vec k|$ is the magnitude of the particle momentum in the rest-frame, $z$ is the cosine of the center-of-mass (c.m.) scattering angle and $P_L(z)$ is the Legendre polynomial.
For a $z$-independent amplitude (\ref{eq37a}):
\begin{equation}
\label{F0}
F_0(k_v)=\frac{1}{16\pi} {\cal F}(M^2_{12}).
\end{equation}

In the given normalization, the scattering amplitude is related to the phase shift by \cite{ck2b}
\begin{equation}\label{F0a}
F_0(k_v)=\frac{\varepsilon_k}{k_v}\exp(i\delta_0)\sin\delta_0= \varepsilon_k f_0(k_v),
\end{equation}
with $\varepsilon_k=\sqrt{k_v^2 + m^2}$. The S-matrix is unitary if the  phase-shift  $\delta_0$ is  real and 
\begin{equation}\label{f0k}
f_0(k_v)=\frac{1}{k_v \cot \delta_0 -ik_v},
\end{equation}
is the standard non-relativistic form of the scattering amplitude. The expansion in powers of $k_v^2$ in the low energy region, gives: 
\begin{equation}\label{1a}
k_v\cot\delta_0= - \frac{1}{a}+\frac{1}{2}r_0k_v^2+\cdots,
\end{equation}
where $a$ is the scattering length and $r_0$ the effective range. The relation to ${\cal F}(M^2_{12})$ is 
\begin{equation}\label{fF}
f_0(k_v)=\frac{1}{16\pi \varepsilon_k}{\cal F}(M^2_{12}),
\end{equation}
and the  two renormalization  conditions to be used in the following, are
based either on fixing
 the bound state pole or the  scattering length.
 The latter condition reads
\begin{equation}\label{ren}
f_0(k_v=0)= \frac{1}{16\pi m}{\cal F}(4m^2)=-a,
\end{equation}
that fixes the scattering amplitude at the continuum branch point.
 
\subsection{Renormalization via bound state pole}\label{SubSec:deriv_F}

One way to calculate ${\cal B}(M^2_{12})$ is to use  the standard Feynman parametrization:
\begin{equation}
\label{Eq:Feynman_param}
\frac{1}{a\,b}=\int_0^1\frac{du}{[ua+(1-u)b]^2},
\end{equation}
with $a=k^2-m^2+i\epsilon$, $b=(k-P)^2-m^2+i\epsilon$, and then compute the 4D integral in the Euclidean space. However, for $M^2_{12}\geq 4m^2$, the integrand of this integral becomes singular and this method is not so convenient. 

Therefore, to calculate the amplitude (especially for  $M_{12}^2>4m^2$) we use  the initial integral (\ref{eq38}) (written in the c.m.~frame $\vec{P}=\vec{0}$) and start by performing the integration over  $k_0$ by residues, i.e.
\begin{equation}\label{eq41}
\begin{aligned}
{\cal B}(M^2_{12})&=-\int\frac{dk_0d^3k}{(2\pi)^4}\frac{1}{(k_0^2-k_v^2-m^2+i\epsilon)}\\
&\times\frac{1}{[(k_0-M_{12})^2-k_v^2-m^2+i\epsilon]}=\\
&2\pi i(res_1(M_{12})+res_2(M_{12})).
\end{aligned}
\end{equation}
Here $res_{1,2}$ are the residues of the integrand in one of the two poles  in the upper half plane of the complex variable $k_0$. The positions of the poles are
$$
k^{(1)}_0=-\varepsilon_k+ i\epsilon,\quad k^{(2)}_0=M_{12}-\varepsilon_k+ i\epsilon,
$$
and the corresponding residues are given by 
\begin{small}
\begin{eqnarray}
res_1(M^2_{12})&=&\int_0^\Lambda\frac {k_v^2 dk_v}{(2\pi)^3}\frac{1}{\varepsilon_k}\frac{1}{[(\varepsilon_k+M_{12})^2-\varepsilon_k^2+i\epsilon]},
\label{res1}
\\
res_2(M^2_{12})&=&\int_0^\Lambda\frac {k_v^2 dk_v}{(2\pi)^3}\frac{1}{\varepsilon_k}\frac{1}{[M_{12}(M_{12}-2\varepsilon_k)+i\epsilon]},
\label{res2}
\end{eqnarray}
\end{small}
where   the integrals are regularized by  the momentum cut-off $\Lambda$.
If $-\infty<M_{12}^2<4m^2$, the integrals  (\ref{res1})  and (\ref{res2}) are non-singular ones. Contrary to this, if $M_{12}>2m$, the second residue is represented as a sum of two contributions: the principal value of the integral over $k_v$ and the delta-function contribution, i.e.
 \begin{equation}
\begin{aligned}
\; &res_2(M^2_{12})= res_{2a}(M^2_{12})+res_{2b}(M^2_{12}) \\
&=
\;PV\int_0^\Lambda\frac {k_v^2 dk_v}{(2\pi)^3}\frac{1}{\varepsilon_k}\frac{1}{M_{12}(M_{12}-2\varepsilon_k)}\\
&+\;\int_0^\Lambda\frac {k_v^2 dk_v}{(2\pi)^3}\frac{1}{\varepsilon_k}(-i\pi)\delta[M_{12}(M_{12}-2\varepsilon_k)]\\
&= res_{2a}(M^2_{12})+\frac{1}{2\pi i}\frac{y''}{16\pi} ,
\end{aligned}
\label{eq42}
\end{equation} 
where the delta function is integrated out by taking  $\Lambda \to \infty$  and $y''$ is defined below, in Eq. (\ref{eqy}).

As the contribution $res_{2a}(M^2_{12})$ in \eqref{eq42} is divergent in the ultraviolet limit, it is necessary to perform a regularization process. By renormalizing we can express the bare parameters (in this work, the coupling constant $\lambda$) via observables (usually, in the field theory, via a ``physical" coupling constant). 
From the condition that the two-body system has a bound state with the mass $M_{2}$ and  the 
amplitude (\ref{eq37a}) has a pole at $M_{12}=M_{2}$, one finds for the coupling constant $\lambda$
\begin{small}
\begin{equation}
(i \lambda)^{-1}={\cal B}(M^2_{2})=2\pi i(res_1(M^2_{2})+res_2(M^2_{2})).
\label{lambda2}
\end{equation}
\end{small}
The denominator in (\ref{eq37a}) then becomes
\begin{multline}
i[(i \lambda)^{-1}-{\cal B}(M^2_{12})]=i[{\cal B}(M^2_{2B})-{\cal B}(M^2_{12})]\\
= i\,PV\int_0^{\infty}\frac{(M_{2}^2-M^2_{12}) }
{32\pi^2\,\varepsilon_k \left[k_v^2-\left(\frac{1}{4}M^2_{12}-m^2\right)\right] }\\
\times\frac{k_v^2 dk_v}{\left[k_v^2+\left(m^2-\frac{1}{4}M^2_{2}\right)\right]} -\frac{y''}{16\pi}.
\label{den1}
\end{multline}
In Eq.~\eqref{den1}, the principal value (PV) integral takes into account the singularity at  $k_v=\sqrt{\frac{1}{4}M^2_{12}-m^2}$ and 
the limit of $\Lambda\to \infty$ is taken since the integral is ultraviolet finite.

Now the integral (\ref{den1}), (in which the bare coupling constant $\lambda$ is expressed via the  two-body bound state mass $M_2$) 
 is finite  and its  calculation in different domains of the variable $M_{12}$ results for $\mathcal{F}(M^2_{12}) $ in:
\begin{eqnarray}
&&\mbox{({\it i}) If $-\infty < M_{12}^2\leq 0$\quad $(1\geq y\geq 0)$, then:}
\nonumber\\
 &&\mathcal{F}(M^2_{12})=
 \left[{\frac{1}{16\pi^2 y}\log\frac{1+y}{1-y}-\frac{\arctan y'_{M_{2}}}{8\pi^2 y'_{M_{2}}}}\right]^{-1}.
 \label{Eq:F_amp1}\\
&&\mbox{({\it ii}) If   $0\leq M_{12}^2 \leq 4m^2 \quad (0 \leq y' < \infty)$,  then:} 
\nonumber\\
&& \mathcal{F}(M^2_{12})=\left[{\frac{\arctan y'}{8\pi^2 y'}-\frac{\arctan y'_{M_{2}}}{8\pi^2 y'_{M_{2}}}}\right]^{-1}.
 \label{Eq:F_amp2}\\
&&\mbox{({\it iii}) If   $ 4m^2\leq  M_{12}^2<\infty \quad (0 \leq  y'' \leq 1)$,  then:}
\nonumber\\ 
&& \mathcal{F}(M^2_{12})=\left[{\frac{y''}{16\pi^2}\log\frac{1+y''}{1-y''}-\frac{\arctan y'_{M_{2}}}{8\pi^2 y'_{M_{2}}}-i\frac{y''}{16\pi}}\right]^{-1}.
 \label{Eq:F_amp3}
\end{eqnarray}
Here $ y'_{M_{2}}=\frac{M_{2}}{\sqrt{4m^2-M_{2}^2}} $
and
\begin{equation} \label{eqy}
\begin{aligned}
&y=\frac{\sqrt{-M_{12}^2}}{\sqrt{4m^2-M_{12}^2}}, 
\quad y'=\frac{M_{12}}{\sqrt{4m^2-M_{12}^2}} ,\\
&y''=\frac{\sqrt{M_{12}^2-4m^2}}{M_{12}}.
\end{aligned}
\end{equation}
We have not yet  introduced the scattering length $a$ in  $\mathcal{F}(M^2_{12})$ instead of the two-body bound state mass,  
which will allow to generalize the scattering amplitude for the case where no bound state exists. This will be discussed in  detail in the next subsection.
 It should be also  noticed that above the threshold  $M_{12}>2m$,  the amplitude obtains in the denominator an imaginary part $-i\frac{y''}{16\pi}.$
 
\subsection{Renormalization via scattering length}

 The  two-body scattering amplitude \eqref{Eq:F_amp1}-\eqref{Eq:F_amp3} was obtained by fixing a bound state  pole at $M_{12}=M_{2}$. However, it can happen that the two-body bound state is absent, hence, the two-body scattering amplitude has no  any bound state pole.  Besides that, the three-body bound state may   exist in absence of the two-body one.
In this situation a different condition will be used: the requirement that the scattering amplitude at zero 
energy is equal to $-a$, where $a$ is the two-body scattering length. 
The (non-renormalized) two-body amplitude  ${\cal F}(M^2_{12})$ still has the form of Eq. (\ref{eq37a}).  Its argument can be written as $M^2_{12}=4\varepsilon_k^2$.  
By using (\ref{eq37a}) and (\ref{ren}) we obtain
\begin{equation}
\frac{1}{16\pi m} \frac{1}{i[(i \lambda)^{-1}-{\cal B}(4m^2)]}=-a,
\end{equation}
and therefore
\begin{equation}\label{lambda_a}
(i\lambda)^{-1}={\cal B}(4m^2)-\frac{1}{16i\pi m a}.
\end{equation}

The two-body amplitude is then given by
\begin{equation}\label{Fa}
{\cal F}(M_{12}^2)=\frac{1}{i\Bigl[{\cal B}(4m^2)-{\cal B}(M^2_{12})\Bigr]-\frac{1}{16\pi m a}}.
\end{equation}

The two-body scattering amplitude is obtained after substituting $2m$ for $M_2$ 
in Eq.~(\ref{den1}), and in   the different regions of $M_{12}^2$ it reads:
\begin{eqnarray}
&&\mbox{({\it i}) If $-\infty < M_{12}^2\leq 0$\quad $(1\geq y\geq 0)$, then:}
\nonumber\\
&&{\cal F}(M^2_{12})=16\pi
\left[{{\frac{1}{\pi y}\log \frac{1+y}
{1-y} -\frac{1}{  m a}}}\right]^{-1}.
\label{eqfa1}
\\
&&\mbox{({\it ii}) If   $0\leq M_{12}^2 \leq 4m^2 \quad (0 \leq y' < \infty)$,  then:} 
 \nonumber\\
 &&{\cal F}(M^2_{12})= 16\pi
\left[{{\frac{2}{ \pi}\frac{\arctan y'}{y'} -\frac{1}{ m a}}}\right]^{-1}.
\label{eqfa2}
\\
&&\mbox{}
\nonumber\\
&&\mbox{({\it iii}) If   $ 4m^2\leq  M_{12}^2<\infty \quad (0 \leq  y'' \leq 1)$,  then:}
\nonumber\\
&&{\cal F}(M^2_{12})=16\pi
\left[{\frac{y''}{\pi}\log \frac{1+y''}{1-y''}
 -\frac{1}{ m a}-i y''}\right]^{-1}.
\label{eqfa3}
\end{eqnarray}
For negative scattering length $a$ this amplitude  
has no poles.

The two-body amplitude can  be written from Eq.~\eqref{eqfa3} in terms of the c.m.~frame momentum  as:
\begin{equation}\label{Fb}
{\cal F}(M_{12}^2)=
16\pi\left[{\frac{k_v}{\varepsilon_k\pi}\log\frac{\varepsilon_k+k_v}{\varepsilon_k-k_v}-\frac{1}{m a}-i\frac{k_v}{\varepsilon_k}}\right]^{-1},
\end{equation}
and then:
\begin{equation}
k_v\cot\delta_0=\frac{k_v}{\pi}\log\frac{\varepsilon_k+k_v}{\varepsilon_k-k_v}-\frac{\varepsilon_k}{m a},
\end{equation}
which is real showing the  unitarity of the model amplitude.  The power expansion for small $k_v$ and comparison with \eqref{1a} allows to identify the effective range as 
\begin{equation}
 r_0=\frac{4}{m\pi}-\frac{1}{m^2a} \, ,
\end{equation}
which is  determined by the terms $\propto \: 1/m,\;1/m^2$, reflecting the relativistic origin of the model. 

In the case when  the bound state exists,  one finds that
\begin{equation}
\label{ar0}
\begin{aligned}
a&=\frac{\pi y'_{M_{2}}}{2m\arctan(y'_{M_{2}})}, \\
 r_0&=\frac{2\Bigl[2y'_{M_{2}}-\arctan(y'_{M_{2}})\Bigr]}{\pi m\,y'_{M_{2}}},
\end{aligned}
\end{equation}
and Eqs.~\eqref{eqfa1}-\eqref{eqfa3} then coincide with \eqref{Eq:F_amp1}-\eqref{Eq:F_amp3}. Furthermore, for small binding energy $B\ll m$ the variable $y'_{M_{2}}$ increases as $y'_{M_{2}}\sim \sqrt{m/B}$.
 The scattering length $a$ also increases for $B\to 0$ with
\begin{equation}\label{ar1}
 a\to\frac{1}{\sqrt{mB}}\quad \text{and} \quad r_0\to\frac{4}{m\pi},
\end{equation}
whereas the effective radius $r_0$ tends to a constant.


\section{Three-body Bethe-Salpeter equation}\label{Sec:BSE}

The solution of the zero-range three-body BS equation for three identical spinless particles 
using the Faddeev decomposition of the full BS amplitude can be reduced to the solution of one single integral 
equation for the spectator  vertex function $v_M(q,p)$ (external propagators are excluded). In the zero-range interaction case $v_M(q,p)$ depends upon both the total momentum $p$ and on the four-momentum of the spectator particle $q$. The equation reads \cite{tobias1}:
\begin{equation}
\begin{aligned}
v_M(q,p)&=2i\mathcal{F}(M^2_{12})\int \frac{d^4 k}{(2\pi)^4}\frac{i}{[k^2-m^2+i\epsilon]}\\
&\times\frac{i}{[(p-q-k)^2-m^2+i\epsilon]}v_M(k,p).
\label{Eq:BSE}
\end{aligned}
\end{equation}
Notice that the momentum of the spectator particle, $q$, determines the effective mass of the two-boson subsystem, $M_{12}$, due to the four-momentum conservation (see below). Therefore,  the vertex function does not depend on other momenta besides $q$ and  the total four-momentum $p$. 
The other two components of the integral equation~\eqref{Eq:BSE} can be easily obtained through the cyclic permutation of the momentum of the constituent particles.
The full BS amplitude in Minkowski space is recovered by multiplying the vertex function by
the three external propagators and summing up the components, i.e.
\begin{equation}\label{Eq:Phi_M}
\begin{aligned}
&i\Phi_M(k_1, k_2, k_3; p)=\\
&i^3\frac{v_M(k_1)+v_M(k_2)+v_M(k_3)}{(k^2_1-m^2+i\epsilon)(k^2_2-m^2+i\epsilon)(k^2_3-m^2+i\epsilon)},
\end{aligned}
\end{equation}
where $v_M(k)\equiv v_M(k,p)$ (to simplify our notation) and the four-momenta obey the relation
\begin{equation}
k_1+k_2+k_3=p.
\end{equation}
The relativistic two-body zero-range scattering amplitude $\mathcal{F}(M^2_{12})$   in (\ref{Eq:BSE}) was derived in Sec.~\ref{Sec:two_body_amp} and, being renormalized via scattering length,  is given by equations  \eqref{eqfa1}, \eqref{eqfa2} and \eqref{eqfa3}.
Its argument $M^2_{12}$ is expressed via three-body momenta as $M_{12}^2=(p-q)^2$. One major simplification in Eq.~(\ref{Eq:BSE}) happens due to the fact that the amplitude $\mathcal{F}(M^2_{12})$ does not depend on the loop integration variable $k$ in the zero-range case. Due to that the two-body amplitude factors out in the integral equation, what does not happen for a finite-range interaction kernel like the one-boson exchange or the cross-ladder one.

 Notice that in Refs.~\cite{tobias1,ck3b} the regime $M_{12}^2 > 4m^2$ of ${\cal F}(M^2_{12})$ (\ref{eqfa3}) was not presented, due to the range of the variables considered in those works.
The amplitude in terms of the bound state mass, as presented in Refs.~\cite{tobias1,ck3b}, is given in Eq.~\eqref{Eq:F_amp2} (i.e.~in the physical domain, $0\leq M_{12} \leq 2m$).

Such a link is very important to understand the range covered by the results obtained 
previously, in Refs.~\cite{tobias1,ck3b}, by considering only the situation where the two-body state is bound (i.e. $a>0$)  producing a real $M_2$ through Eqs.~\eqref{ar0}), and 
the full support covered by equations  \eqref{eqfa1}, \eqref{eqfa2} and \eqref{eqfa3}, including also virtual two-body bound states (i.e. $a \in \mathbb{R}$).
In other words, as mentioned, in the region for which $a<0$ the amplitude $\mathcal{F}(M^2_{12})$ has no pole in the physical domain and, therefore, the two-body bound state does not exist. The three-body system can still be formed though, as a Borromean bound state. 

The goal now is to solve the scalar three-body BS equation \eqref{Eq:BSE}, derived in Ref.~\cite{tobias1}, for  the lowest angular momentum bound state, 
with zero-range interaction, fully in Minkowski space and retaining 
implicitly the Fock-space composition beyond the valence truncation.  
The adopted method is the direct integration of the singularities of the four-dimensional integral equation, developed recently 
for the two-body BS equation in Ref.~\cite{ck2b}. The method does not rely on any 
ansatz, as e.g.~the Nakanishi integral representation  used for the solution of the two-body equation in Ref.~\cite{FrePRD14}. No three-dimensional reduction of the covariant 4D equation, as the one 
done by performing the projection onto the LF plane, is adopted.
Part of what is exposed here was published in Ref.~\cite{Ydrefors19_3b_Mink}. Further comparisons with  results obtained in the Euclidean space calculations will be provided to test the reliability of the method. 

One interesting example of a calculation within the approach used here is the  electromagnetic transition form 
factor~\cite{Carbonell:2015awa}, which quantifies the breakup of a two-body bound state. This highly complex calculation was performed through the direct integration method in Minkowski space, using as inputs the solutions, obtained by the same method~\cite{ck2b}, of the scattering and bound state BS equations. 
The transition form factor, including the final state 
interaction, was calculated in the whole kinematical region. It satisfied the non-trivial condition of current conservation explicitly verified numerically.

Equation \eqref{Eq:BSE} is a singular integral equation and solving 
it numerically is a very challenging task. For that reason, the 
equation requires a proper treatment to be rewritten in, at least, a less singular 
form before its numerical solution in the c.m.~frame, $\vec{p}=\vec{0}$. The propagators, containing the strongest 
singularities of the BS equation kernel, are represented in the customary form~\cite{ck2b}
\begin{equation}
\label{Eq:prop_repr}
\begin{aligned}
&\frac{1}{k^2-m^2+i\epsilon}=\frac{1}{k_0^2-k^2_v-m^2+i\epsilon}
\\ &=PV\frac{1}{k_0^2-\varepsilon_k^2}
-\frac{i\pi}{2\varepsilon_k}[\delta(k_0-\varepsilon_k)+\delta(k_0+\varepsilon_k)],
\end{aligned}
\end{equation} 
where $PV$ denotes the principal value.  The terms like $PV\int\ldots\frac{dk_0}{k_0^2-\varepsilon_k^2}$ contain the singularities at $k_0=\pm \varepsilon_k$ which  are removed by subtracting integrals from the equation, with appropriate coefficients, in such a way that the final equation is not affected. For that, the following identities are used
\begin{equation}
PV \int_{-\infty}^0 \frac{dk_0}{k^2_0-\varepsilon_k^2}=PV \int_{0}^{\infty} \frac{dk_0}{k^2_0-\varepsilon_k^2}=0.
\end{equation}

The second propagator in Eq.~\eqref{Eq:BSE} can be integrated over the angles analytically. Denoting $z=\cos\left(\frac{\vec{k}\cdot\vec{q}}{k_v q_v}\right)$ and recalling that $d^3k=k_v^2dk_v\,dz\,d\varphi$, one can write that
\begin{eqnarray}\label{Eq:PI}
&&\Pi(q_0,q_v, k_0, k_v)   =\int\frac{idzd\varphi}{[(p-q-k)^2-m^2+i\epsilon]} \nonumber \\
&&=\frac{i \pi}{q_v k_v} \left\{ \log \left| \frac{(\eta+1)}{(\eta-1)} \right| - i \pi I(\eta)   \right\},
\end{eqnarray}
with
\begin{equation}\label{Eq:eta}
I(\eta)=\left\{
\begin{array}{lcrcl}
1  & {\rm if} & \mid\eta\mid &\leq& 1 \cr
0  & {\rm if} & \mid\eta\mid &> & 1
\end{array}\right.,
\end{equation}
and
\begin{equation}
\eta =   \frac{(M_3 - q_0 - k_0)^2 - k_v^2 - q_v^2 - m^2}{2q_v k_v},
\end{equation}
where $M_3$ denotes the bound state mass of the three-body system.
The BS equation \eqref{Eq:BSE} turns, after integration over the angles, into an integral equation with the kernel \eqref{Eq:PI}, that is still singular.  However, these singularities  are weakened by integration and become logarithmic ones and discontinuities, that will be treated numerically.

Once the propagators are expressed as in Eq.~(\ref{Eq:prop_repr}), the principal value singularities are subtracted and the angular integrations are performed, Eq. \eqref{Eq:BSE} acquires the following form:
\begin{widetext}
\begin{eqnarray}
v_M(q_0, q_v)  
&=& \frac{\mathcal{F}(M^2_{12})}{(2\pi)^4} \int_0^{\infty} k^2_v dk_v\left\lbrace   \frac{2\pi i}{2 \varepsilon_k}  \left[\Pi(q_0, q_v; \varepsilon_k, k_v) v_M(\varepsilon_k, k_v)  + \Pi(q_0, q_v; -\varepsilon_k, k_v)  v_M(-\varepsilon_k, k_v)\right]\right. 
\cr
   &-&2 \int^0_{-\infty} dk_0 \left[ \frac{ \Pi(q_0, q_v; k_0, k_v) v_M(k_0, k_v)  - \Pi(q_0, q_v; -\varepsilon_k, k_v)  v_M(-\varepsilon_k, k_v)}{  {k}_0^2-\varepsilon_k^2 }\right]  \cr
&-&  \left.   2 \int_0^{\infty} dk_0 \left[ \frac{ \Pi(q_0, q_v; k_0, k_v) v_M(k_0, k_v)  - \Pi(q_0, q_v; \varepsilon_k, k_v)  v_M(\varepsilon_k, k_v)}{  {k}_0^2-\varepsilon_k^2 }\right]\right\rbrace.
                   \label{Eq:v}
\end{eqnarray}
\end{widetext}
This equation has now, besides the unknown analytical behavior of $v_M(q_0,q_v)$ that will be discovered numerically, only weak singularities and discontinuities, but 
unlike (\ref{Eq:BSE}) the singularities in $k_0=\pm\varepsilon_k$ no longer exist.

The logarithmic singularities of the kernel \eqref{Eq:PI}, $\Pi(q_0,q_v,k_0,k_v)$, at $\eta=\pm 1$ can be  found for fixed values of $q_0$, $q_v$ and $k_v$, what makes the numerical treatment in $k_0$ easier. Their positions with respect to the variable $k_0$ are
\begin{eqnarray}
k_0&=&(M_3-q_0)+\sqrt{m^2+(k_v\pm q_v)^2},
\nonumber\\
k_0&=&(M_3-q_0)-\sqrt{m^2+(k_v\pm q_v)^2}.
\label{Eq:sing_k0}
\end{eqnarray}

Analogously, the position of the singularities can be found for the variable $k_v$, so that the integration over this variable can be optimized numerically. The positions of the singularities of $\Pi(q_0,q_v,\pm\varepsilon_k,k_v)$ as a function of $k_v$ are given by
\begin{equation}
k_v=\frac{\pm \sqrt{M_{12}^2(M_{12}^2+q_v^2)(M_{12}^2-4m^2)}\pm q_v M_{12}^2}{2M_{12}^2},
\label{Eq:sing_k}
\end{equation}
where $M_{12}^2=(M_3-q_0)^2-q^2_v$. The expression under the square root is non-negative if
\begin{equation}
M_{12}^2 \geq 4m^2 \quad \mbox{or} \quad M_{12}^2 \leq 0,
\end{equation}
and, therefore, for existing real singularities in $k_v$ one needs to ensure one of the following conditions for $q_0$: $q_0<M_3-\sqrt{q^2_v+4m^2}$ or $M_3-q_v<q_0<M_3+q_v$ or  $q_0>M_3+\sqrt{q^2_v+4m^2}$.
This means that the branching points that need to be considered while fixing the mesh numerically to separate the regions with and without singularities in $k_v$ are
\begin{equation}
\label{Eq:peaks}
\begin{aligned}
q^{(1)}_0=&M_3-\sqrt{q^2_v+4m^2}, \\
q^{(2)}_0=&M_3-q_v, \\
q^{(3)}_0=&M_3+q_v, \\
q^{(4)}_0=&M_3+\sqrt{q^2_v+4m^2},
\end{aligned}
\end{equation}
with $q^{(1)}_0<q^{(2)}_0< q^{(3)}_0< q^{(4)}_0$. As it can be seen from Eqs.~\eqref{Eq:F_amp1}-\eqref{Eq:F_amp3}, these branching points are also present in the two-body amplitude $\mathcal{F}(M_{12}^2)$. For more details on the behavior of the $\mathcal{F}(M_{12}^2)$ amplitude, see Sec.~\ref{Sec:F_num}. 

\section{Relation between the BS amplitude and LF wave function}

In this section and in Appendix \ref{App:BSA_LFWF}, we establish the relation between the three-body  BS amplitude and the three-body LF wave function (LFWF). It generalizes to the three-body system  the relation (3.57) or (3.58) from Ref.~\cite{cdkm} for the two-scalar system and can be easily generalized to the arbitrary $n$-body case. 

The three-body BS amplitude is defined analogously to the two-body one, namely:
\begin{equation} \label{Phi0_1}
\Phi_M(x_1,x_2,x_3;p)=\langle 0 \left| T\Bigl(\varphi (x_1)\varphi (x_2)\varphi (x_3)\Bigr)\right| p\rangle\, .
\end{equation}

As is shown in Appendix \ref{App:BSA_LFWF}, the three-body LFWF can be related to the BS amplitude through 
\begin{equation} \label{bs7b_1}
\begin{aligned}
\psi (&\vec{k}_{1\perp},\xi_1;\vec{k}_{2\perp},\xi_2;\vec{k}_{3\perp},\xi_3) = \frac{(p^+)^2 }{\sqrt{2\pi}}\, \xi_1 \xi_2 \xi_3
\\
&\times \int dk^-_{1}\, dk^-_{2}
 \Phi_M(k_1,k_2,k_3;p),
\end{aligned}
\end{equation}
where $\xi_1+\xi_2+\xi_3=1$ and the plus and minus momentum components are given by
$p^{\pm}=p^0\pm p^3$, with analogous definitions for the other momenta.
  
Similarly, one has for the two-body case that
\begin{equation}
\label{bs8_1}
\begin{aligned}
\psi (\vec{k}_{1\perp},&\xi_1;\vec{k}_{2\perp},\xi_2)=\\
 &=\frac{p^+}{\sqrt{2\pi}} \, \xi_1 \xi_2 \int dk^-_{1} \Phi_M(k_1,k_2;p).
\end{aligned}
\end{equation}
By introducing the relative variable $k=\frac{1}{2}(k_1-k_2)$, $\Phi_M\equiv\Phi_M(k,p)$, Eq. (\ref{bs8_1}) can be written on the form:
\begin{small}
\begin{equation} \label{bs9_1}
\psi (\vec{k}_{\perp},\xi)=\frac{p^+}{\sqrt{2\pi}} \xi(1- \xi)\int dk^{-} \Phi_M(\vec{k}_{\perp},k^+\hspace{-.1cm},k^-;p).
\end{equation}
\end{small}
It differs from Eq.~(3.57) of Ref.~\cite{cdkm} by the degrees of $\pi$ due to the presence of the factor $(2\pi)^{-3/2}$ in Eq.~(3.53)  of Ref.~\cite{cdkm}  which we did not introduce above.

Noteworthy that the support of the function  
$\psi (\vec{k}_{\perp},\xi)$ in the variable $\xi$, or of the integral 
$$\int dk^{-} \Phi_M(\vec{k}_{\perp},k^+,k^-;\,p),$$ is $0<\xi<1$, as it should be.  
This follows from the fact that $\Phi_M$ is not an arbitrary function, but it is defined, in the coordinate space, 
by $$\Phi_M(x_1,x_2;p)=
\langle 0 \left| T\Bigl(\varphi (x_1)\varphi (x_2)\Bigr)\right| p\rangle\, ,$$
analogously to the three-boson BS amplitude defined in Eq.~(\ref{Phi0_1}). 
This support is also automatically obtained if one represents the BS amplitude in the Nakanishi form (see Ref.~\cite{bs1}, Appendix D). 
The same conclusion is also valid for the three-body case, when using Eq.~(\ref{bs7b_1}).

\section{Non-relativistic limit}
In this section, the non-relativistic limits of the three-body Euclidean BS and valence LF equations, i.e.~Eqs.~(7) and (10) of Ref.~\cite{ey3b}, are considered. The Euclidean BS equation will be first analysed. Representing the three-body mass  $M_3$ as $M_3=3m-B_3$, with $B_3$ denoting the three-body binding energy, and truncating the denominator  in Eq.~(7) of Ref.~\cite{ey3b}, and the terms in the fraction of the argument of the $\log$ in Eq.~(8) of the aforementioned reference, to the leading terms of momenta and the binding energies, one gets
\begin{equation}
\label{K}
\begin{aligned}
K&= \frac {\Pi_E(q_4,q_v,k_4,k_v)}{\left(k_4-\frac{i}{3}M_3\right)^2+k_v^2+m^2}=\\
&\frac{\frac{1}{2}\log\frac{\left(k_4+q_4+\frac{i}{3}M_3\right)^2+(q_v+k_v)^2+m^2}{\left(k_4+q_4+\frac{i}{3}M_3\right)^2+(q_v-k_v)^2+m^2}}
{\left(k_4-\frac{i}{3}M_3\right)^2+k_v^2+m^2}
\\
\Rightarrow\;K_{nr}& \approx\frac{\frac{1}{2}\log\frac{\frac{2}{3}B_3+\frac{(k_v+q_v)^2}{2m}+i(k_4+q_4)}
 {\frac{2}{3}B_3+\frac{(k_v-q_v)^2}{2m}+i(k_4+q_4)}}{2m\left(\frac{1}{3}B_3-ik_4\right)}.
 \end{aligned}
\end{equation}
At the  first glance, one could neglect the terms $\frac{(k_v\pm q_v)^2}{2m}$  in comparison to $(k_4+q_4)$, however, this would result in $K_{nr}\equiv 0$, and therefore it is necessary to keep them.

Following Ref.~\cite{tobias1}, one can write
$E_2$ through $M_{12}^2=(2m-E_2)^2$,
  the two-body bound state mass as $M_2=2m-B_2$ and introduce them
  in the two-body amplitude ${\cal F}(M_{12}^2)$ in the physical domain ($0\leq M_{12}^2\leq 4m^2$) (see Eq.~(\ref{Eq:F_amp2}))  to elaborate the non-relativistic limit. 
  In such  case, $m\to\infty$, and the ${\cal F} (M_{12}^2)$ amplitude becomes 
\begin{equation}\label{Fnr}
{\cal F}(M^2_{12})=\frac{16\pi\sqrt{m}}{\sqrt{E_2}-\sqrt{B_2}}, 
\end{equation}
or, alternatively,
\begin{equation}\label{Fnr1}
{\cal F}(-M^2_{12})=\frac{16\pi\sqrt{m}}{\sqrt{2m-\sqrt{-M^2_{12}}}-\sqrt{B_2}}.
\end{equation}
Since  ${M}_{12}^2= \left(\frac{2}{3}p-iq_4\right)^2-q^2_v = -(\frac{2}{3}iM_3+q_4)^2-q^2_v$ in the limit $m\to\infty$ one gets
\begin{equation}
\begin{aligned}
E_2&=2m-\sqrt{-\left[\frac{2}{3}i(3m-B_3)+q_4\right]^2-q^2_v}\\
&\approx  \frac{2}{3}B_3+iq_4+\frac{q_v^2}{4m}.
\end{aligned}
\end{equation}

Substituting it in (\ref{Fnr}), one finds for the scattering amplitude
\begin{equation}\label{Fnr2}
{\cal F}(M^2_{12})=\frac{16\pi\sqrt{m}}{\sqrt{ \frac{2}{3}B_3+iq_4+\frac{q_v^2}{4m}}-\sqrt{B_2}}. 
\end{equation}
After these manipulations, Eq.~(7) of Ref.~\cite{ey3b} obtains the form
\begin{equation}\label{eqnr}
\begin{aligned}
\tilde{v}'_E(&q_4,q_v)= \frac{1}{\pi^2\sqrt{m}}\frac{1}{\sqrt{\frac{2}{3}B_3+iq_4+\frac{q_v^2}{4m}}-\sqrt{B_2}} \\
&\times\int_0^{\Lambda} dk_v \int_{-\infty}^{\infty}\frac{dk_4}{\left(\frac{1}{3}B_3-ik_4\right)}\\
&\times\log\frac{\frac{2}{3}B_3+\frac{(k_v+q_v)^2}{2m}+i(k_4+q_4)}
 {\frac{2}{3}B_3+\frac{(k_v-q_v)^2}{2m}+i(k_4+q_4)}\;\tilde{v}'_E(k_4,k_v),
\end{aligned}
\end{equation}
where it was introduced a cutoff $\Lambda$ to prevent the Thomas collapse~\cite{thomas}. In order to obtain the time independent equation, the integration over $k_4$ needs to be performed. Since this is a  lengthy derivation, it will not be done here explicitly.

For the three-body LF equation given by Eq.~(10) of Ref.~\cite{ey3b}, the non-relativistic limit, obtained by following the same steps as before, reads
\begin{equation}\label{LFnr}
\begin{aligned}
\Gamma_{nr}(&\vec{q})=\frac{1}{\pi^2m^{3/2}}\frac{1}{\sqrt{E_2}-\sqrt{B_2}}\\
&\times\int\frac{\Gamma_{nr}(\vec{k})d^3k}{{B_3+\frac{q_v^2}{2m}+\frac{k_v^2}{2m}+
\frac{(\vec{q}+\vec{k})^2}{2m}}},
\end{aligned}
\end{equation}
where
\begin{equation}
\begin{aligned}
E_2&=2m-M_{12}\approx B_3+\frac{3}{4}\frac{q_v^2}{m}. 
\end{aligned}
\end{equation}
Here the factor $\frac{1}{\sqrt{E_2}-\sqrt{B_2}}$ is originating from the two-body amplitude (\ref{Fnr}) when $m\to\infty$.

Eq.~(\ref{LFnr}) is the same as Eq.~(18) of Ref.~\cite{tobias1}. This equation is known as the Skornyakov-Ter-Martirosyan equation~\cite{STMeq}.
The non-relativistic equation  can be also written in the form
\begin{equation}\label{nr18}
\begin{aligned}
\Gamma_{nr}(&\vec{q})=\frac{1}{\pi^2\sqrt{m}}
\frac{1}{\sqrt{B_3+\frac{3}{4}\frac{q_v^2}{m}}-\sqrt{B_{2}}}\\
&\times\int \frac{\Gamma_{nr}(\vec{k})d^3k}
{k_v^2+\vec{k}\cd\vec{q}+q_v^2+mB_3},
\end{aligned}
\end{equation}
and, for the s-wave, after integrating over the angles, it reads
\begin{equation}\label{nr20}
\begin{aligned}
\Gamma&_{nr}(q_v)=\frac{2}{\pi\sqrt{m}}
\frac{1}{\sqrt{B_3+\frac{3}{4}\frac{q_v^2}{m}}-\sqrt{B_2}}\\
&\times\int_0^{\Lambda}  \log\left(\frac{k_v^2+k_vq_v+q_v^2+mB_3}{k_v^2-k_vq_v+q_v^2+mB_3}\right)
\Gamma_{nr}(k_v)\frac{k_vdk_v}{q_v}.
\end{aligned}
\end{equation}
The above equation, like (\ref{eqnr}), and in contrast to \eqref{Eq:BSE}, requires a cutoff in order to allow a physical solution  avoiding the Thomas collapse~\cite{thomas}. 

\section{Transverse amplitudes}\label{Sec:TA}

The vertex function $v(q_0, q_v)$ is fundamentally dependent on the metric (Euclidean or Minkowski one) adopted to define the integral equation. 
The transverse amplitude is, instead, an useful quantity for comparison between calculations performed in Euclidean and Minkowski spaces. 
Furthermore, it gives information on the valence wave function integrated in the longitudinal momenta  in the present model, where 
an infinite number of Fock-components are taken into account implicitly by the Bethe-Salpeter framework.  The rich structure of the
three-boson bound state  transverse amplitude is investigated numerically in Sec.~\ref{Sec:Res_TA}, as it is a mean to exploit the  double parton momentum dependence 
of the valence wave function, and also gives access to the dynamical correlation between the constituents.
 
 The derivation of the expressions for the Minkowski transverse amplitude is presented below in Sec.~\ref{Ssec:transv_mink}. The final amplitude, computed with the BS amplitude obtained from the solution of the BS equation in Minkowski space (\ref{Eq:v}), is expected to coincide with the one defined in Euclidean space. The expressions for the latter one will be derived in Sec.~\ref{Ssec:transv_eucl}.

\subsection{Minkowski space}\label{Ssec:transv_mink}

As mentioned, the  BS amplitude $\Phi_M$ can be written in terms of the three vertex components by introducing the external propagators. It is given above by Eq.~(\ref{Eq:Phi_M}).

The transverse amplitude can be defined via $\Phi_M$ as
\begin{equation}\label{Eq:L_Mink}
\begin{aligned}
L(&\vec{k}_{1\perp}, \vec{k}_{2\perp})=
\\&L_1(\vec{k}_{1{\perp}}, \vec{k}_{2{\perp}})+L_2(\vec{k}_{1{\perp}}, \vec{k}_{2{\perp}})+L_3(\vec{k}_{1{\perp}}, \vec{k}_{2{\perp}})=\\
& \int_{-\infty}^{\infty}dk_{10} \int_{-\infty}^{\infty} dk_{1z} \int_{-\infty}^{\infty} dk_{20} \int_{-\infty}^{\infty} dk_{2z}\\
&\times i\Phi_M(k_{10}, k_{1z}, k_{20}, k_{2z}; \vec{k}_{1\perp}, \vec{k}_{2\perp}).
\end{aligned}
\end{equation}

However, in the three identical boson case, only one of its Faddeev components $L_i$ is enough for the comparison with the transverse amplitude derived from the Euclidean BS solution. The first Faddeev component is given by
\begin{equation}
\begin{aligned}
L_1(\vec{k}_{1\perp}, \vec{k}_{2\perp})&=i\int_{-\infty}^{\infty}dk_{10}\int_{-\infty}^{\infty}dk_{1z}\;\frac{v_M(k_{10}, k_{1v})}{k_1^2-m_1^2+i\epsilon}\\
&\times\chi(k_{10}, k_{1z}; \vec{k}_{1\perp}, \vec{k}_{2\perp}),
\end{aligned}
\end{equation}
with
\begin{equation}
\label{Eq:chi}
\begin{aligned}
\chi(& k_{10}, k_{1z};\vec{k}_{1\perp}, \vec{k}_{2\perp})=\\
&i^2\int\frac{d^2k_2}{(k^2_2-m_2^2+i\epsilon)[(p'-k_2)^2-m_3^2+i\epsilon]},
\end{aligned}
\end{equation}
where the following quantities enter:
$k_i=(k_{i0}, k_{iz})$ and $d^2k_i= dk_{i0} dk_{iz}$ with $i=1,2$. Moreover, 
\begin{small}
\begin{equation}\label{m1m2}
m_2^2=m^2+|\vec{k}_{2\perp}|^2,  \quad m_3^2=m^2+(\vec{p}_{\perp}-\vec{k}_{1\perp}-\vec{k}_{2\perp})^2,
\end{equation}
\end{small}
and $p'=(p'_0, p'_z)=p-k_1=(p_0-k_{10},p_z-k_{1z})$.

The two-dimensional integral in \eqref{Eq:chi} can be \\ performed by first introducing the Feynman \\ parametrization \eqref{Eq:Feynman_param}
and then making the transformation $k_2 \rightarrow k_2 + (1-u)p'$. After integration over $k_2$, using a Wick rotation as $k_0=ik_4$, we find
\begin{equation}
\label{Eq:chi_3}
\begin{aligned}
\chi(k_{10}, k_{1z}; &\vec{k}_{1\perp}, \vec{k}_{2\perp}) 
=i^2\int_{0}^1 du\int\frac{d^2k_2}{(k^2_2+D+i\epsilon)^2} \\
&= -\pi i^3\int_{0}^{1}\frac{du}{D+i\epsilon},
\end{aligned}
\end{equation} 
with
\begin{equation}
\label{Eq:den_M}
D = u(1-u)p'^2 - m^2_2u - (1-u)m^2_3.
\end{equation}

The denominator $D$ is zero at
\begin{equation}
\begin{aligned}
\label{upum}
u&_{\mp}=\frac{1}{2p'^2}\left[p'^2-m_2^2+m^2_3\right.\\
&\mp\left.\sqrt{((m_2-m_3)^2-p'^2)((m_2+m_3)^2-p'^2)}\right],
\end{aligned}
\end{equation}
but for $p'^2 < (m_2+m_3)^2,$ the equality $D=0$ is never satisfied in the interval $0<u<1$, so the term $i\epsilon$ can be dropped out in Eq.~\eqref{Eq:chi_3} and the integral over the Feynman parameter $u$ can be performed safely analytically, giving the following
\begin{equation}
\begin{aligned}
\chi(k_{10}&, k_{1z}; \vec{k}_{1\perp}, \vec{k}_{2\perp}) = \frac{\pi i^3}{p'^2(u_- -u_+)}\\
&\times\int_{0}^{1}du\biggl[\frac{1}{u-u_-}-\frac{1}{1-u_+} \biggr]=\\
& -\frac{i\pi}{p'^2(u_{-}-u_+)}[\log(1-u_-)\\
&-\log(-u_-)-\log(-1+u_+)+\log(u_+)],
\end{aligned}
\end{equation}   
with $u_{\pm}$ defined in \eqref{upum}. 

In the situation where $p'^2 > (m_2+m_3)^2$, the zeroes of the denominator, $u_{\pm}$, are placed on the real axis for the interval $u \in [0,1]$. For that reason, one can separate $\chi$ in two terms, analogously to what was done in \eqref{Eq:prop_repr}, i.e.
\begin{equation}
\begin{aligned}
&\chi(k_{10}, k_{1z}; \vec{k}_{1\perp}, \vec{k}_{2\perp}) = \\
&\chi'(k_{10}, k_{1z}; \vec{k}_{1\perp}, \vec{k}_{2\perp}) + \chi''(k_{10}, k_{1z}; \vec{k}_{1\perp}, \vec{k}_{2\perp}),
\end{aligned}
\end{equation}
 where
\begin{eqnarray}
\label{Eq:chip}
&&\chi'(k_{10}, k_{1z}; \vec{k}_{1\perp}, \vec{k}_{2\perp}) =  \frac{\pi i^3}{p'^2(u_- -u_+)}
\nonumber \\
&&\times\biggl[\text{PV}\int_{0}^{1}\frac{du}{u-u_-}- \text{PV}\int_{0}^{1}\frac{du}{u-u_+} \biggr],
\end{eqnarray}
and
\begin{equation}
\begin{aligned}
\chi''(&k_{10}, k_{1z};\vec{k}_{1\perp}, \vec{k}_{2\perp}) = \frac{\pi i^3}{p'^2(u_- -u_+)}\\
& \times\biggl[-i\pi\int_{0}^{1}du \delta(u-u_-)-i\pi\int_{0}^{1}du \delta(u-u_+) \biggr]\\
&=\frac{2\pi^2}{\sqrt{[p'^2-(m_2-m_3)^2][p'^2-(m_2+m_3)^2]}}.
\end{aligned}
\end{equation}

The principal value integrals in Eq.~\eqref{Eq:chip} can be carried out analytically and one obtains for $\chi'$ the following expression
\begin{equation}
\begin{aligned}
&\chi'(k_{10},k_{1z};\vec{k}_{1\perp},\vec{k}_{2\perp})=\\
&i\pi\frac{\log\frac{m^2_2+m^2_3-p'^2-\sqrt{[p'^2-(m_2-m_3)^2][p'^2-(m_2+m_3)^2]}}{m^2_2+m^2_3-p'^2+\sqrt{[p'^2-(m_2-m_3)^2][p'^2-(m_2+m_3)^2]}}}{\sqrt{[p'^2-(m_2-m_3)^2][p'^2-(m_2+m_3)^2]}}.
\end{aligned}
\end{equation}

The contribution $L_1(\vec{k}_{1{\perp}}, \vec{k}_{2{\perp}})$ can subsequently be written in the form
\begin{multline}
L_1(\vec{k}_{1{\perp}}, \vec{k}_{2{\perp}})=\\ 
-i\int_{-\infty}^{\infty}dk_{1z}\Bigg\{  \frac{i\pi}{2\tilde{k}_{10}}\left[\chi(\tilde{k}_{10}, k_{1z}; \vec{k}_{1\perp},\vec{k}_{2\perp})v_M(\tilde{k}_{10},k_{1v})\right.\\ \left.
+\chi(-\tilde{k}_{10}, k_{1z}; \vec{k}_{1\perp},\vec{k}_{2\perp})v_M(-\tilde{k}_{10},k_{1v})\right]\\
-\int_0^{\infty}dk_{10}\left[\frac{\chi(-k_{10},k_{1z};\vec{k}_{1\perp},\vec{k}_{2\perp})v_M(-k_{10},k_{1v})}{k^2_{10}-\tilde{k}^2_{10}}\right.\\
\left.-\frac{\chi(-\tilde{k}_{10},k_{1z};\vec{k}_{1\perp},\vec{k}_{2\perp})v_M(-\tilde{k}_{10},k_{1v})}{k^2_{10}-\tilde{k}^2_{10}}\right]\\
  -\int_0^{\infty}dk_{10}\left[\frac{\chi(k_{10},k_{1z};\vec{k}_{1\perp},\vec{k}_{2\perp})v_M(k_{10},k_{1v})}{k^2_{10}-\tilde{k}^2_{10}}
  \right. \\ \left.
 -\frac{\chi(\tilde{k}_{10},k_{1z};\vec{k}_{1\perp},\vec{k}_{2\perp})v_M(\tilde{k}_{10},k_{1v})}{k^2_{10}-\tilde{k}^2_{10}}\right]\Bigg\}, 
\label{Eq:transverse_final}
\end{multline}
where
\begin{equation}
\tilde{k}_{10}=\sqrt{k^2_{1z}+|\vec{k}_{1\perp}|^2+m^2}.
\end{equation}
Similarly to the treatment of the BS equation in Sec.~\ref{Sec:BSE}, propagators like  $[k^2_1-m^2_1+i\epsilon]^{-1}$ were expressed in the form \eqref{Eq:prop_repr} and  subtractions were made to eliminate the principal value singularities at $k_0=\pm \tilde{k}_{10}$.

It should be noticed that the function $\chi$ in Eq.~\eqref{Eq:transverse_final} has square-root singularities at $p'^2=(m_2\pm m_3)^2$. The functions $\chi(\pm\tilde{k}_{10}, k_{1z};\vec{k}_{1\perp},\vec{k}_{2\perp})$ are thus singular at 
\begin{multline}
\label{kroots}
k_{1z}=\pm\frac{2}{M_3}\sqrt{(M_3+m_1)^2-(m_2+m_3)^2} \\ 
\times \sqrt{(M_3-m_1)^2-(m_2+m_3)^2}.
\end{multline}
Furthermore, for fixed $k_{1z}$, the positions of the singular points of the functions $\chi(-k_{10},k_{1z};\vec{k}_{1\perp},\vec{k}_{2\perp})$ and $\chi(k_{10},k_{1z};\vec{k}_{1\perp},\vec{k}_{2\perp})$ 
are given by
\begin{equation}
k_{10}=-M_3+\sqrt{k^2_{1z}+(m_2+m_3)^2},
\end{equation}
and
\begin{equation}
k_{10}=M_3\pm \sqrt{k_{1z}^2+(m_2+m_3)^2},
\end{equation}
respectively. In this case only the singular points located on the positive $k_0$ axis need to be considered (see Eq.~\eqref{Eq:transverse_final}).
In fact, it turns out that the integrands in Eq.~\eqref{Eq:transverse_final} are symmetric with respect to $k_{1z} \rightarrow -k_{1z}$. Therefore, one  needs to consider only the region where $k_{1z}>0$ and multiply the equation by a factor 2. Furthermore, only the positive solutions of Eq.~\eqref{kroots} are needed.

\subsection{Euclidean space}\label{Ssec:transv_eucl}

The expressions for the transverse amplitudes in the Euclidean space, presented in Ref.~\cite{ey3b}, will be derived in detail in this section and in Appendix \ref{App:transv_eucl_kernel}.
As mentioned in that paper, the following change of variables in the original equation (\ref{Eq:BSE})  was performed,
\begin{equation}
\label{shift}
k_i = k'_i+\frac{p}{3},\quad (i=1, 2, 3), 
\end{equation}
in order to allow the Wick rotation without crossing any singularities. The primed momenta satisfy the relation
\begin{equation}
k'_1+k'_2+k'_3=0.
\end{equation}
The BS amplitude in Minkowski space can be written as
\begin{equation}
  \label{Eq:phi_M}
\begin{aligned}
&i\tilde{\Phi}_M(k'_1,k'_2,k'_3;M_3)=i^3\frac{\tilde{v}_M(k'_1)+\tilde{v}_M(k'_2)+\tilde{v}_M(k'_3)}{(k'_1+\frac{p}{3})^2-m^2+i\epsilon}\\
&\times\frac{1}{[(k'_2+\frac{p}{3})^2-m^2+i\epsilon][(k'_3+\frac{p}{3})^2-m^2+i\epsilon]}=\\
&i^3\frac{\tilde{v}_M(k'_1)+\tilde{v}_M(k'_2)+\tilde{v}_M(-k'_1-k'_2)}{[(k'_1+\frac{p}{3})^2-m^2+i\epsilon][(k'_2+\frac{p}{3})^2-m^2+i\epsilon]}\\
&\times\frac{1}{(k'_1+k'_2-\frac{p}{3})^2-m^2+i\epsilon},
\end{aligned}
\end{equation}
where
\begin{equation}
\tilde{\Phi}_M(k'_1,k'_2,k'_3;p)=\Phi_M\bigl(k'_1+\frac{p}{3},k'_2+\frac{p}{3},k'_3+\frac{p}{3};p\bigr),
\end{equation}
and
\begin{equation}
  \label{Eq:vtilde}
\tilde{v}_M(k'_i) = v_M\bigl(k'_i+\frac{p}{3}\bigr).
\end{equation}

Now, in new (shifted) integration variable one can perform the Wick rotation in the equation (\ref{Eq:BSE}) and transform this equation into the Euclidean space. For the full Euclidean BS amplitude one gets:
\begin{equation}\label{Eq:phi_E}
\begin{aligned}
i\tilde{\Phi}_E(&k'_{14}, k'_{1z}, \vec{k}'_{1\perp};k'_{24}, k'_{2z}, \vec{k}'_{2\perp})=\\
-&i^3\frac{\tilde{v}_E(k'_{14},k'_{1v})+\tilde{v}_E(k'_{24},k'_{2v})+\tilde{v}_E(k'_{34},k'_{3v})}{(k'_{14}-i\frac{M_3}{3})^2+k'^2_{1z}+m^2_1}\\
&\times\frac{1}{(k'_{24}-i\frac{M_3}{3})^2+k'^2_{2z}+m^2_2}\\
&\times \frac{1}{(k'_{14}+k'_{24}+i\frac{M_3}{3})^2+(k'_{1z}+k'_{2z})^2+m^2_3},
\end{aligned}
\end{equation}
where 
\begin{equation}\label{m1m2m3}
\begin{aligned}
k'_{iv} &= \sqrt{|\vec{k}'_{i\perp}|^2+k'^2_{iz}}, \\
m^2_i &= |\vec{k}'_{i\perp}|^2 + m^2, \qquad (i=1,2,3) \\
\vec{k}'_{3\perp} &= -(\vec{k}'_{1\perp}+\vec{k}'_{2\perp}).
\end{aligned}
\end{equation}

The full Euclidean transverse amplitude, corresponding to the Minkowski one given by \eqref{Eq:L_Mink}, reads
\begin{equation}\label{Eq:L_E}
\begin{aligned}
L(&\vec{k}'_{1\perp}, \vec{k}'_{2\perp})=
\\&L_1(\vec{k}'_{1{\perp}}, \vec{k}'_{2{\perp}})+L_2(\vec{k}'_{1{\perp}}, \vec{k}'_{2{\perp}})+L_3(\vec{k}'_{1{\perp}}, \vec{k}'_{2{\perp}})=\\
& -\int_{-\infty}^{\infty}dk'_{14} \int_{-\infty}^{\infty} dk'_{1z} \int_{-\infty}^{\infty} dk'_{24} \int_{-\infty}^{\infty} dk'_{2z}\\
&\times i\tilde{\Phi}_E(k'_{14},k'_{1z},k'_{24},k'_{2z}; \vec{k}'_{1\perp}, \vec{k}'_{2\perp}).
\end{aligned}
\end{equation}

By insertion of Eq.~\eqref{Eq:phi_E} in \eqref{Eq:L_E}, it is found that one of the contributions to the transverse amplitude is given by
\begin{equation}
  \label{Eq:L1_E_final}
\begin{aligned}
L_1(\vec{k}'_{1\perp}, &\vec{k}'_{2\perp})= -\int_{-\infty}^{\infty} dk'_{1z} \int_{-\infty}^{\infty}dk'_{14}\\
&\times\chi(k'_{14},k'_{1z}; \vec{k}'_{1\perp}, \vec{k}'_{2\perp})\tilde{v}(k'_{1v},k'_{14})\\
& \times \frac{i}{(k'_{14}-i\frac{M_3}{3})^2+k'^2_{1z}+m^2_1},
\end{aligned}
\end{equation}
where the function $\chi$ is derived in Appendix \ref{App:transv_eucl_kernel} and reads
\begin{equation}
\label{Eq:chi_E_2}
\begin{aligned}
\chi(k'_{14},&k'_{1z}; \vec{k}'_{1\perp}, \vec{k}'_{2\perp}) = -\pi\int_{0}^1 \frac{du}{A}. 
\end{aligned}
\end{equation}
Here the denominator $A$ is given by
\begin{equation}
A = au^2+bu+c,
\end{equation}
with
\begin{equation}
\label{abc}
\begin{aligned}
a&=-k'^2_{1z}-\Bigl(k'_{14}+\frac{2}{3}iM_3\Bigr)^2,\\
b&=k'^2_{1z}+\Bigl(k'_{14}+\frac{2}{3}iM_3\Bigr)^2+m^2_2-m^2_3,\\
c&=m^2_3,
\end{aligned}
\end{equation}
where the $m_i$'s are defined by Eqs.~(\ref{m1m2}) and (\ref{m1m2m3}).

\section{Wick rotation in the three-body BS equation}\label{Sec:wick_rotation}

  It was shown in Ref.~\cite{ey3b} that the three-body BS equation \eqref{Eq:BSE}, after the introduction of the shifted variables \eqref{shift} allows the Wick rotation without crossing any singularities. The validity of this rotation in the complex plane is a key for the equivalence between the Minkowski- and Euclidean-space transverse amplitudes. Therefore, we outline in this section some of the main points related to the Wick rotation of  \eqref{Eq:BSE}.
 
The BS equation \eqref{Eq:BSE} enclosing the shifted variables takes the form
\begin{equation}
  \label{Eq:BSE_shifted}
  \begin{aligned}
  \tilde{v}_M(q',p) = &2i\mathcal{F}(M'^2_{12})\int \frac{d^4 k'}{(2\pi)^4}\frac{i}{[(k' + \frac{p}{3})^2 - m^2 + i\epsilon]} 
  \\ & \times \frac{i}{[(\frac{p}{3} - q' - k')^2 - m^2 +i\epsilon]}\tilde{v}_M(k',p),
  \end{aligned}
\end{equation}
where $M'^2_{12} = (\frac{2}{3}p - q')^2$ and $\tilde{v}_M$ is defined by \eqref{Eq:vtilde}.

In the center-of-mass frame  the pole in the upper half of the $k'_0$ complex plane of the first propagator is located at
\begin{equation}
  \label{Eq:sing_kpp1}
  k'^{(+)}_{01} = -\frac{M_3}{3} - \sqrt{k'^2_v + m^2} + i\epsilon,
\end{equation}
and for the second one the position of the pole is
\begin{equation}
  \label{Eq:sing_kpp2}
  k'^{(+)}_{02} = \eta' - q'_0 + i\epsilon,
\end{equation}
with
\begin{equation}
  \label{Eq:eta_p}
  \eta' = \frac{M_3}{3}-\sqrt{(\vec{k}\,'+\vec{q}\,^\prime)^2 + m^2}.
\end{equation}
Consequently, as seen from  \eqref{Eq:sing_kpp1}, the first propagator does not have any poles in the first quadrant. Moreover, since $M_3 < 3m$ one has that $\eta' < 0$. The Wick rotation of \eqref{Eq:BSE_shifted} can thus be done safely without crossing any singularities.

Contrary to this, for the BS equation in the form \eqref{Eq:BSE}, written in terms of the unshifted variables (i.e.~$k$ and $q$), the second propagator has a pole at
\begin{equation}
  \label{Eq:sing_kp2}
  k^{(+)}_{02} = \eta - q_0 + i\epsilon,
\end{equation}
where
\begin{equation}
  \label{Eq:eta_new}
  \eta = M_3-\sqrt{(\vec{k}+\vec{q})^2 + m^2},
\end{equation}
and if $M_3>m$, there exist  $\vec{k}$ and $\vec{q}$ such that $\eta >0$, and the Wick rotation is thus not permitted.

Each of the three Faddeev components of the transverse amplitute, given by \eqref{Eq:L_Mink}, is like the right-hand side of the BS equation \eqref{Eq:BSE}, an integral over a vertex function times a product of propagators. It is therefore clear that the Wick rotation (after introduction of the shifted variables) also should hold for the transverse amplitude. The propagators entering the definition of the BS amplitude \eqref{Eq:phi_M} have poles  of the form \eqref{Eq:sing_kpp1}, and should therefore not cause any additional problems in this respect. The expected equivalence between the transverse amplitudes computed in Minkowski and Euclidean spaces respectively, is confirmed by  the numerical results to be presented in Sec.~\ref{Sec:Res_TA}. Though, in this paper  the transverse amplitudes, depending on $k_x, k_y$, are compared, we want to stress that the aforementioned arguments are also applicable to the more general amplitudes, not integrated over $k_z$, i.e., depending on $k_x, k_y, k_z$. 

\section{Results and Discussion}\label{Sec:Res}

\subsection{Vertex function in Minkowski space}

The Faddeev component of the vertex function is quantitatively studied in Minkowski space, as it carries the dynamical content of the relativistic three-body model. 
From it the full BS amplitude of the system can be constructed. In Minkowski space it has a nontrivial analytic structure since several branch points given by (\ref{Eq:peaks}) are present  in the
kernel of Eq.~\eqref{Eq:v} , and reflected in the cusps appearing in the vertex function, as will be presented in what follows. Evidently, the Euclidean equation, obtained after the Wick rotation (see Ref.~\cite{ey3b}) does not present in its integration path any singularity as well as the corresponding solution.
Despite of this, both solutions can be compared as we are going to present.  

We solved Eq.~\eqref{Eq:v} adopting a spline decomposition of the vertex function $v_M(q,p)$ used in Ref.~\cite{Ydrefors19_3b_Mink}, see Appendix \ref{Sec:NM}. 
The inputs are the scattering length $a$ and the three-body binding 
energy $B_3$, the same as in the Euclidean calculations performed in Ref.~\cite{ey3b}. In the numerical solution of \eqref{Eq:v}, we multiplied the right-hand side of it by a parameter (eigenvalue) $\lambda$. A consistent solution then corresponds to an eigenvalue of $\lambda=1.0$.

Three results for the eigenvalue are given in Table \ref{Tab:results}, for the following values of the two-body scattering length: $am=-1.280$, $am=-1.500$ and $am=-1.705$.   
The results for the eigenvalue $\lambda$, expected to be real and equal to one, 
present small deviations from the unity and also an imaginary part.

Nevertheless, it is important to mention another potential source of error: cutoffs were introduced to constrain the domains of the variables $q_v$ and $q_0$. It is very difficult to reach a reasonable convergence considering the full domains, as the size of the region where the singularities (given by Eqs.~\eqref{Eq:sing_k0} and \eqref{Eq:sing_k}) appear is enlarged along the axes. Moreover, the asymptotic regions start at larger momenta.

The actual values used to truncate the variables were $q^{\text{max}}_v/m=6.0$ and $q^{\text{max}}_0/m=13.0$, for the two smallest binding energies, or $q^{\text{max}}_0/m=15.0$, for the case where $B_3/m=1.001$.
Regardless, the convergence was reached within  about 10\% for the worst case.
On the other hand, in the Euclidean calculations it is possible to take into account the whole range of the involved variables $q_v$, $q_4$, $k_v$ and $k_4$ through a mapping procedure.
The fact that in the Minkowski approach cutoffs were applied while the whole domains were used in the Euclidean calculations makes the results not fully comparable. This might be one of the reasons why in $\lambda$ small non-zero imaginary parts appear and for the deviations from $1$ obtained in the real part.

\begin{table}[!htbp]
\centering
\begin{tabular}{c c c}
\toprule
$B_3/m$   & $a m$    &  $\lambda$ \\
\midrule
$0.006$ & $-1.280$  & $0.999-0.054i$ \\
$0.395$ & $-1.500$   & $1.000+0.002i$ \\
$1.001$ & $-1.705$ &  $0.997+0.106i$\\
\bottomrule
\end{tabular}
\caption{Eigenvalues of the three-body ground state  for three scattering lengths, $a$, computed in \cite{Ydrefors19_3b_Mink} by using the Euclidean three-body binding energies. \label{Tab:results}}
\end{table}

The rest of this section will  be devoted to a detailed study of the representative case with $B/m = 0.395$ and $am = -1.5$. Though, important to point out that  many of the stated conclusions are valid also for the other cases in Table \ref{Tab:results}.

\begin{figure}[!htbp]
\centering
\includegraphics[scale=0.35]{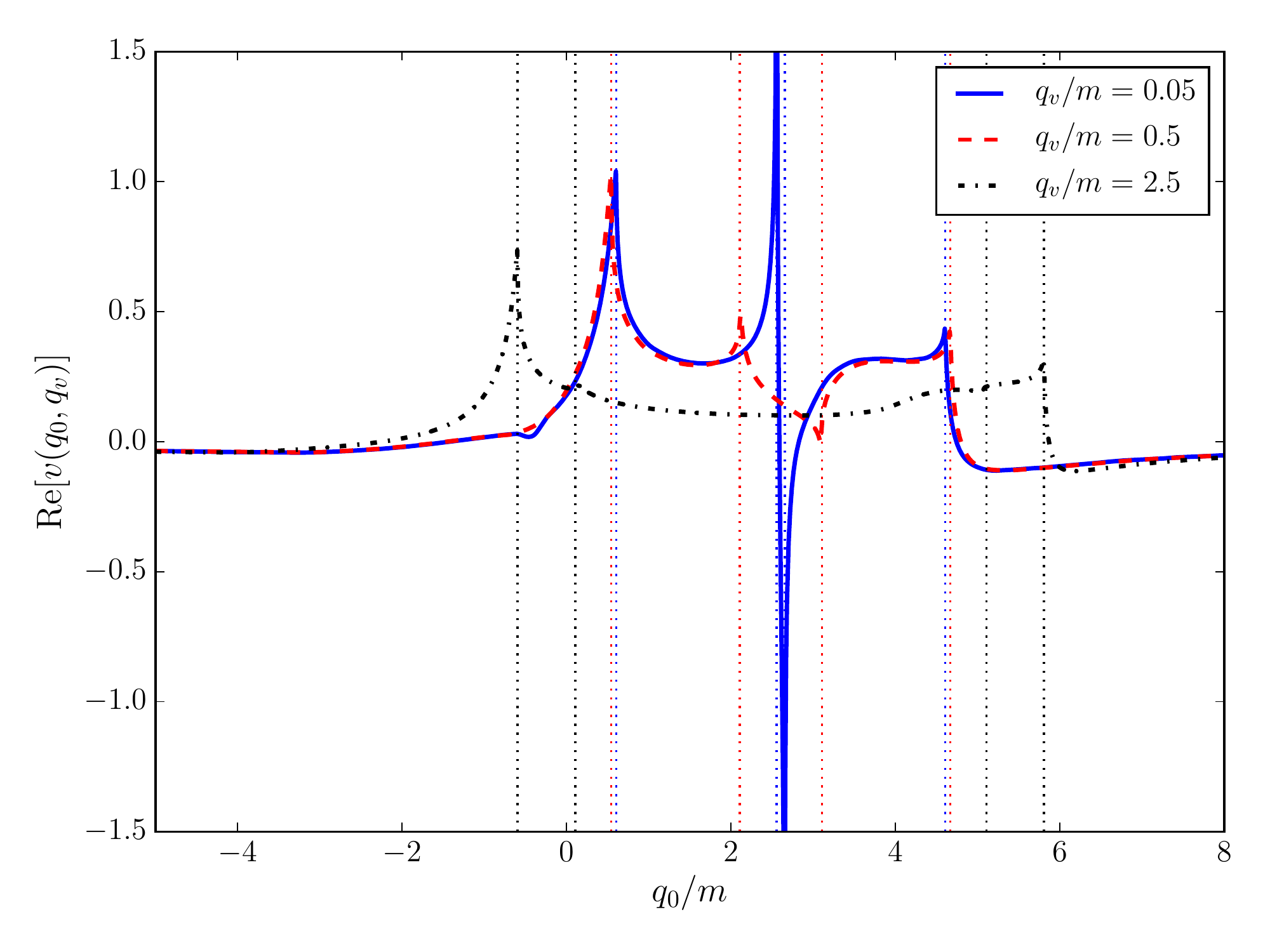}
\includegraphics[scale=0.35]{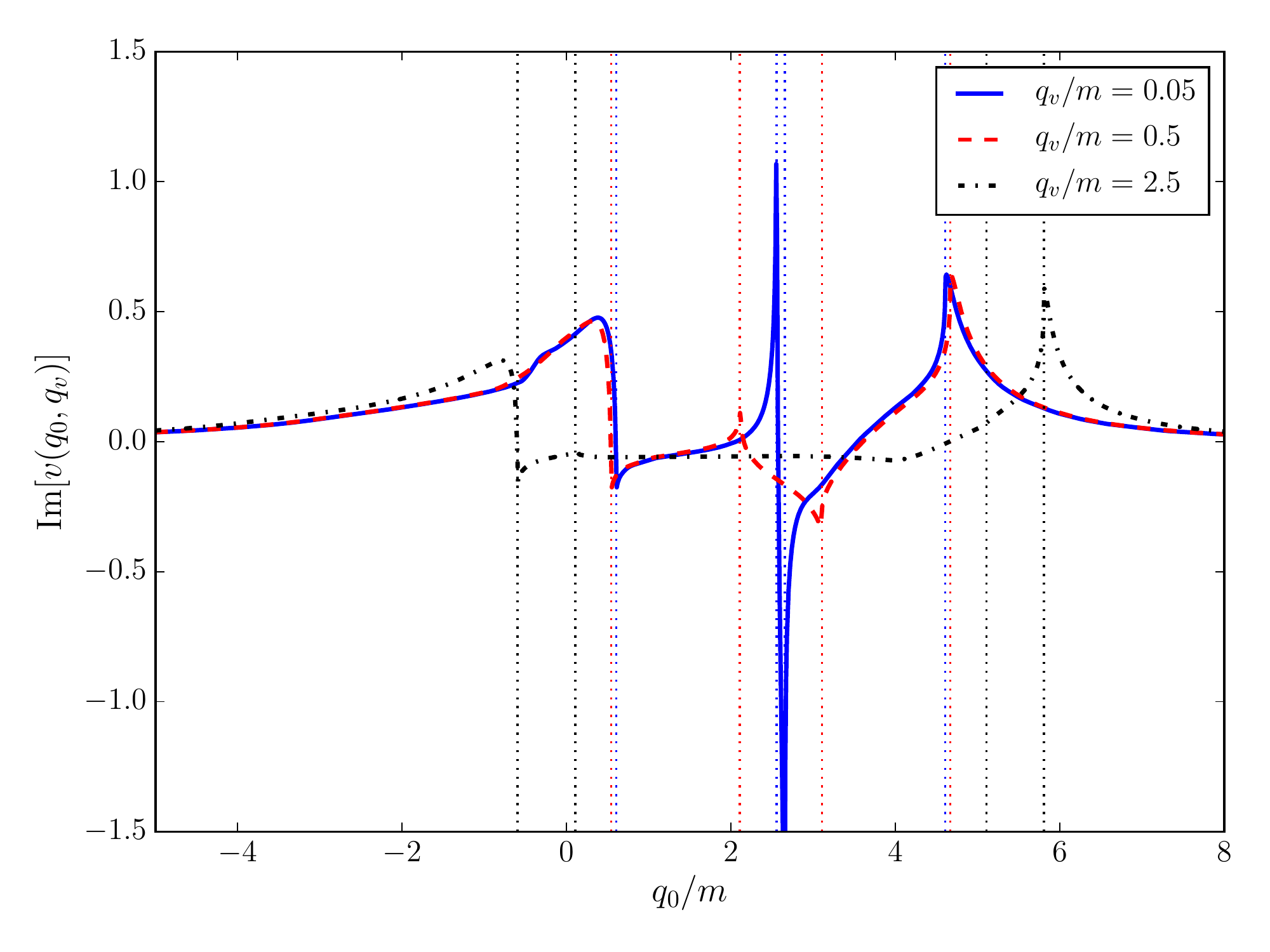} 
\caption{Real (upper panel) and imaginary (lower panel) parts of the vertex function, $v_M(q_0, q_v)$ with respect to  $q_0$.  For each value of $q_v$ the analytical positions of the peaks, given in Eq.~\eqref{Eq:peaks}, are shown with the vertical dotted lines.\label{Fig:v}}
\end{figure}

In Fig.~\ref{Fig:v} it is shown the calculated real and imaginary parts of the vertex function $v_M(q_0, q_v)$ versus $q_0$ for three fixed values of $q_v$. For all three cases there is a quite good agreement between the  numerical peak positions and the analytical ones, given by Eq.~\eqref{Eq:peaks}. Though, worth to mention that, due to the scale of the figure, some of the peaks for the case $q_v/m = 2.5$  are not clearly visible.  The peaks in Fig.~\ref{Fig:v} appear as branching points of the kernel $\Pi(q_0,q_v,\pm \varepsilon_k, k_v)$, defining its singularities, as discussed in Sec.~\ref{Sec:BSE}. 
Interestingly, the aforementioned positions correspond to $M^2_{12}=0$ and $M^2_{12}=4\,m^2$, which give the branching points of the two-body scattering amplitude $\mathcal{F}(M_{12}^2)$. In Fig.~\ref{Fig:v} it is seen that for small values of $q_v$ a singularity appears at $q_0 \approx M_3$. The distance between the external peaks, corresponding to $M^2_{12}=4\,m^2$, is equal to  $2\sqrt{q_v^2+4m^2}$, an increasing function with respect to $q_v$. This fact makes things more complicated from the numerical point of view, as for large values of $q_v$ a very wide region of $q_0$ has to be covered. This imposes the need of cutoffs for the variables.

\begin{figure}[!htbp]
\centering
\includegraphics[scale=0.3]{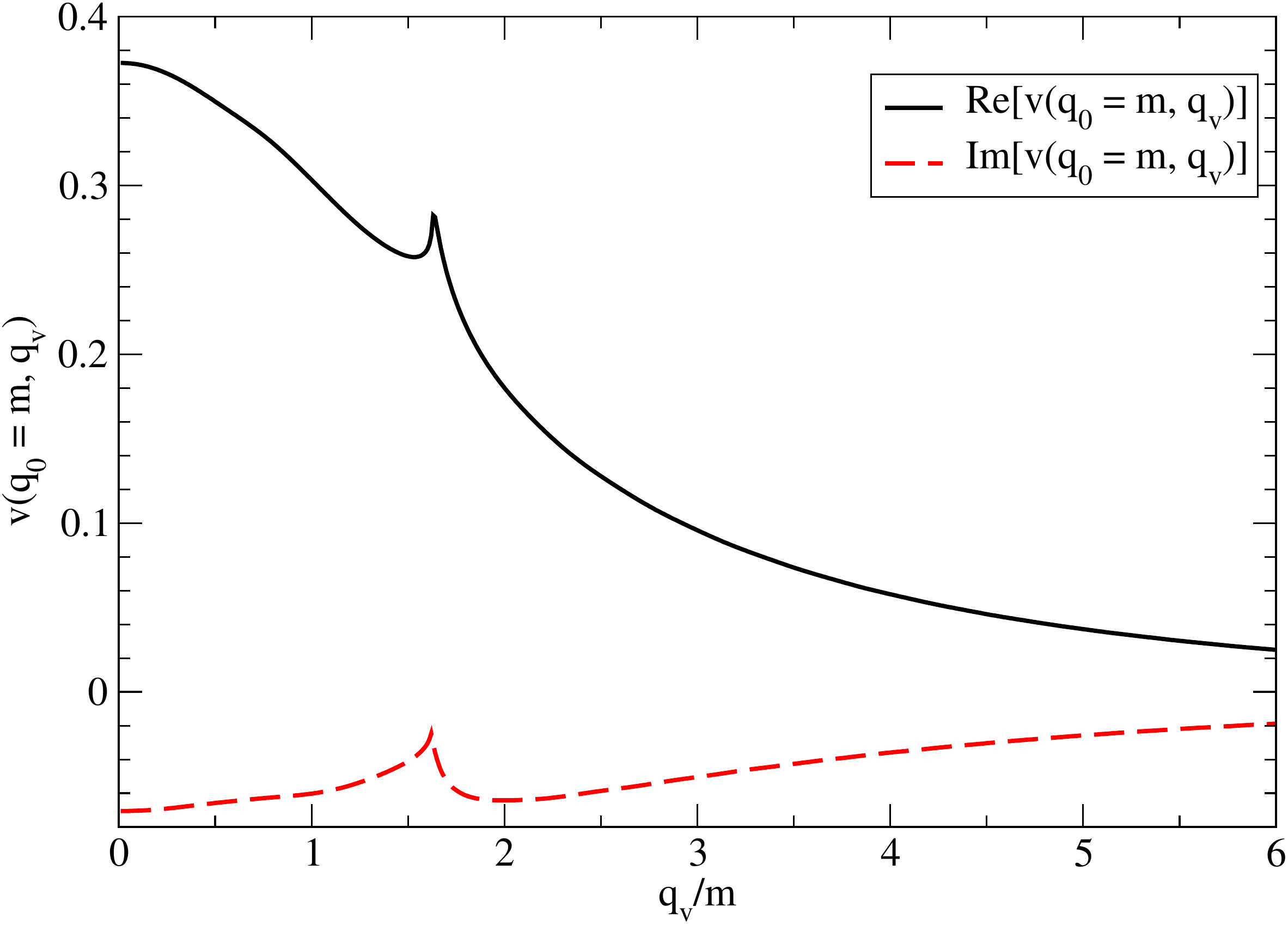} 
\caption{Real and imaginary parts of the vertex function, $v_M(q_0 = m, q_v)$ with respect to  $q_v$.  \label{Fig:v_q}}
\end{figure}

Similarly, in Fig.~\ref{Fig:v_q} we present the  real and imaginary parts of $v_M(q_0, q_v)$ with respect to $q_v$, for $q_0 = m$. In the figure is seen a peak (both in the real and imaginary part) which corresponds to the branching point $q_v= M_3-q_0$. However, for the other points given by Eq.~\eqref{Eq:peaks}, no solution exists such that $q_0= m $ and $q_v \geqslant 0$. It can be seen in the figure that the amplitude asymptotically goes to zero for large values of $q_v$, as expected.

\begin{figure}[!htbp]
\centering
\includegraphics[scale=0.6]{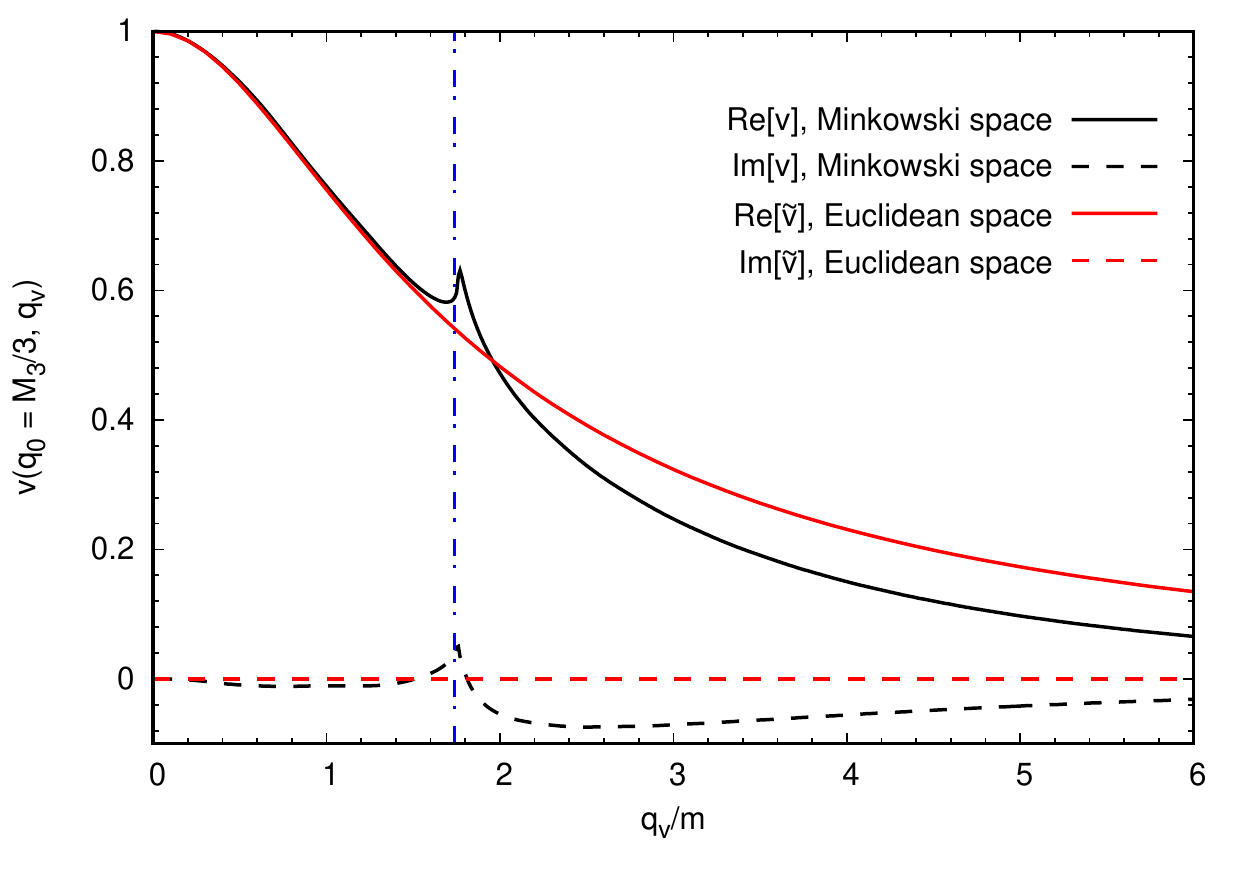} 
\caption{Real and imaginary parts of the Minkowski space vertex function, $v_M(q_0 = M_3/3, q_v)$ versus  $q_v$, compared with the corresponding Euclidean results for $\tilde{v}_E(q'_4,q'_v)$. \label{Fig:v_q_Mink_Eucl}} 
\end{figure}

Furthermore, in Fig.~\ref{Fig:v_q_Mink_Eucl}, we compare for $q_0 = M_3/3$ the calculated real and imaginary parts of $v_M(q_0, q_v)$ versus $q_v$ with the corresponding Euclidean results for $\tilde{v}_E(q',p)= v_E(q' + p/3, p)$, i.e.~$q'_4=0$.  It is seen that the results for both real and imaginary parts are practically the same for small values of $q_v$. However, the  Minkowskian amplitude has a peak at the branching point $q_v=2M_3/3$ and the amplitudes also differ significantly at larger values of $q_v$,  a difference coming from the contribution of the cut necessary to be taken into account to match the limit of $q'_4\to 0$ with the Minkowski point. For the sake of comparison both amplitudes are normalized to 1 for $q'_4=0$ and $q_v=0$.  

\subsection{Transverse amplitude}\label{Sec:Res_TA}

Although the transverse amplitude  has  a different meaning with respect to the  standard distribution amplitude (see Refs.~\cite{LB1980,BPP1988}), it shares in common  with the distribution amplitude  the valence wave function 
content of the full LFWF in Fock space. However, the transverse amplitude has a distinctive feature with respect to the distribution amplitude as it  can  be obtained 
as well from the BS amplitude in Euclidean space, making it a useful quantity to compare with the corresponding Minkowski space quantity. The demonstration of equivalence between the transverse amplitude obtained in Minkowski and Euclidean spaces has been given in Ref.~\cite{GutiPLB2016} resorting to the Nakanishi integral representation of the BS amplitude
 \cite{Nakanishi63, Nakanishi69} in a two-boson system.   Furthermore, as discussed in Sec.~\ref{Sec:wick_rotation}, the Wick rotation of the BS equation \eqref{Eq:BSE} can  be performed without crossing any singularities and should also hold for the transverse amplitude.  Consequently,   the transverse amplitudes computed in Minkowski and Euclidean spaces should agree with each other also for the three-body system, which is  confirmed by the results presented in this section.
 
 In what follows we will exploit the structure of the transverse amplitude associated with one Faddeev component. 
 The reader has to keep in mind that the full transverse amplitude from the three-body wave function is a coherent sum of the three components. However, to expose the consequences of our dynamical assumptions it is simpler to look individually to each Faddeev component, as the sum of $L_i(|\vec{k}_{i\perp}|,|\vec{k}_{j\perp}|,\theta_{ij})$ where $\theta_{ij}$ denotes the angle between $\vec{k}_{i\perp}$ and $\vec{k}_{j\perp}$, will present a much more complex  3D landscape as a function of the two independent transverse momenta.

\begin{figure}[!htbp]
\includegraphics[scale=0.29]{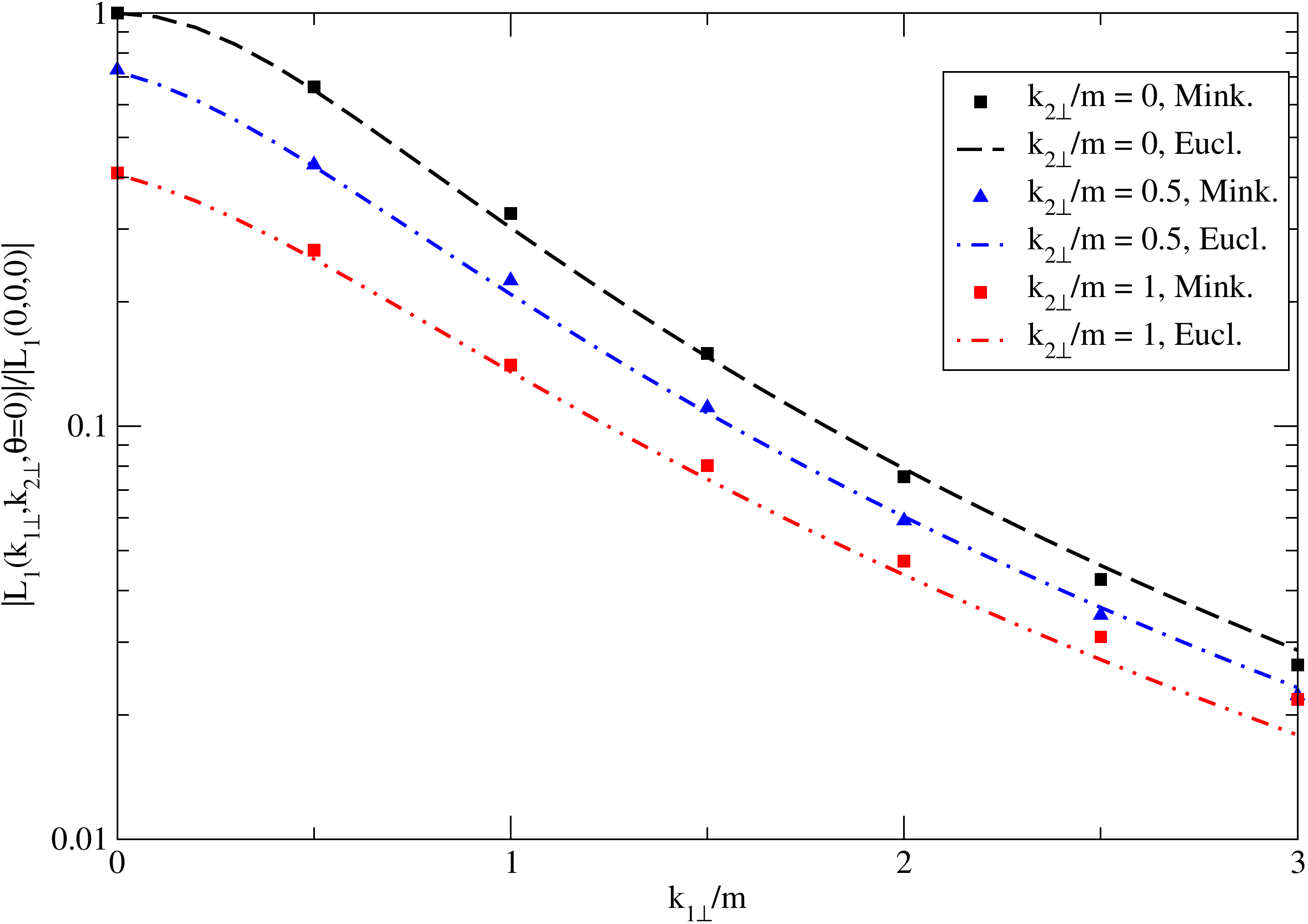}  
\caption{Transverse amplitude modulus,  $|L_1|$ for $\theta=0$  versus $|\vec{k}_{1\perp}|/m$ for   $|\vec{k}_{2\perp}|/m=0.0, 0.5, 1.0$,  obtained in Minkowski space compared with the one calculated
  in Euclidean space. \label{Fig:transverse}}
\end{figure}

 In  Fig.~\ref{Fig:transverse} it is displayed the modulus of the contribution $L_1(|\vec{k}_{1\perp}|,|\vec{k}_{2\perp}|,\theta)$ to the transverse amplitude versus $|\vec{k}_{1\perp}|$, calculated from Eq.~\eqref{Eq:transverse_final}. 
  It is also shown, for comparison, the corresponding Euclidean results calculated through Eq.~\eqref{Eq:L1_E_final}. It is seen  that the Minkowski and Euclidean results are in fair agreement with each other. The non-smooth behavior of the BS solution in Minkowski space, shown in Fig.~\ref{Fig:v}, is washed out and makes the agreement between the Euclidean and Minkowski space transverse amplitudes even more remarkable. 
\begin{figure}[!htbp]
\includegraphics[scale=0.3]{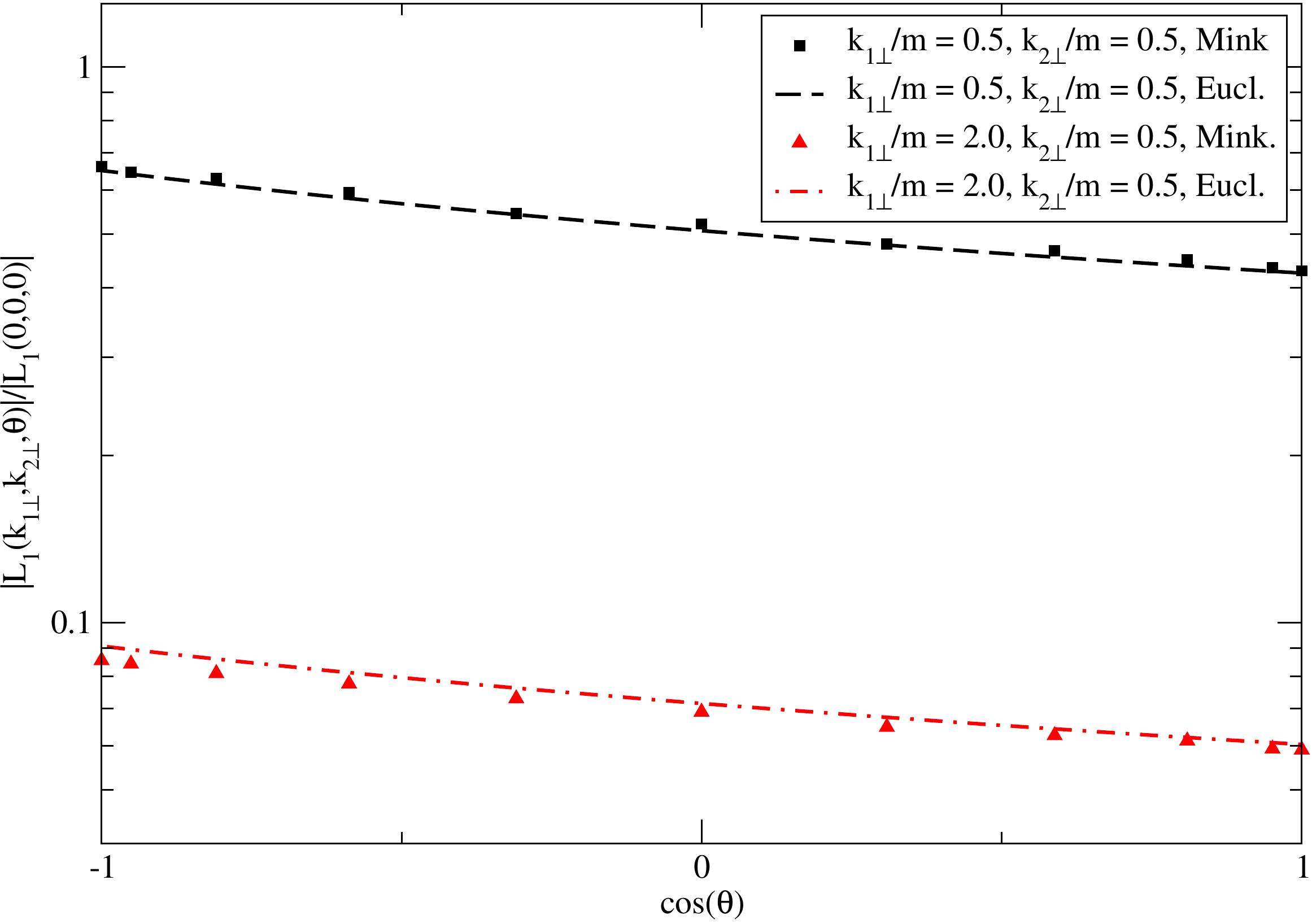}  
\caption{Transverse amplitude modulus, $|L_1|$ as a function of
   $\cos(\theta)$  for $(|\vec{k}_{1\perp}|/m,|\vec{k}_{2\perp}|/m)$ $=(0.5, 0.5), (2.0, 0.5)$, obtained in Minkowski space 
  compared with the one computed in Euclidean space. \label{Fig:transverse_theta}}
\end{figure}

The dependence of the modulus of the transverse amplitude  $|L_1|$ on the angle $\theta$ between $\vec{k}_{1\perp}$ and $\vec{k}_{2\perp}$ is also displayed, in Fig.~\ref{Fig:transverse_theta}.  The modulus of $L_1$ is a slowly decreasing function with respect to $\cos\theta$. As seen in the figure, a satisfactory agreement is again found between the Euclidean (lines) and Minkowski (symbols) calculations. Although the interaction is active in the s-wave and the vertex function is dependent only on
the time component and the modulus of the spatial momentum  in the rest-frame, the Faddeev component of the BS amplitude acquired a weak angular dependence, as we illustrated, due to the presence of the individual propagators and momentum conservation, essentially containing both s- and p-wave dependencies. The mixing of higher waves appears as the binding energy increases, which is expected as the bound state mass becomes smaller increasing the sensitivity of the denominator in Eq.~(\ref{Eq:chi_E_2}) to variations of the relative 
angle, as it is transparent in the case of the transverse amplitude in Euclidean space.

\begin{figure}[!htbp]
\includegraphics[scale=0.5]{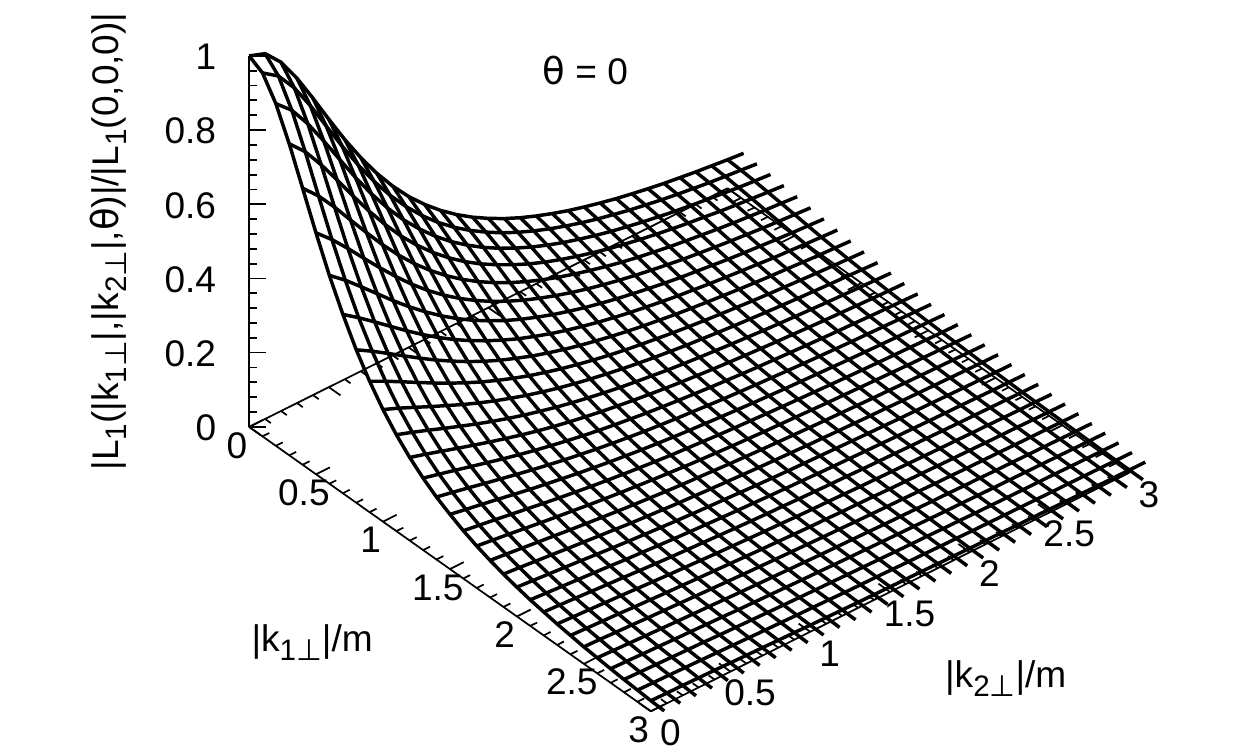}
\includegraphics[scale=0.5]{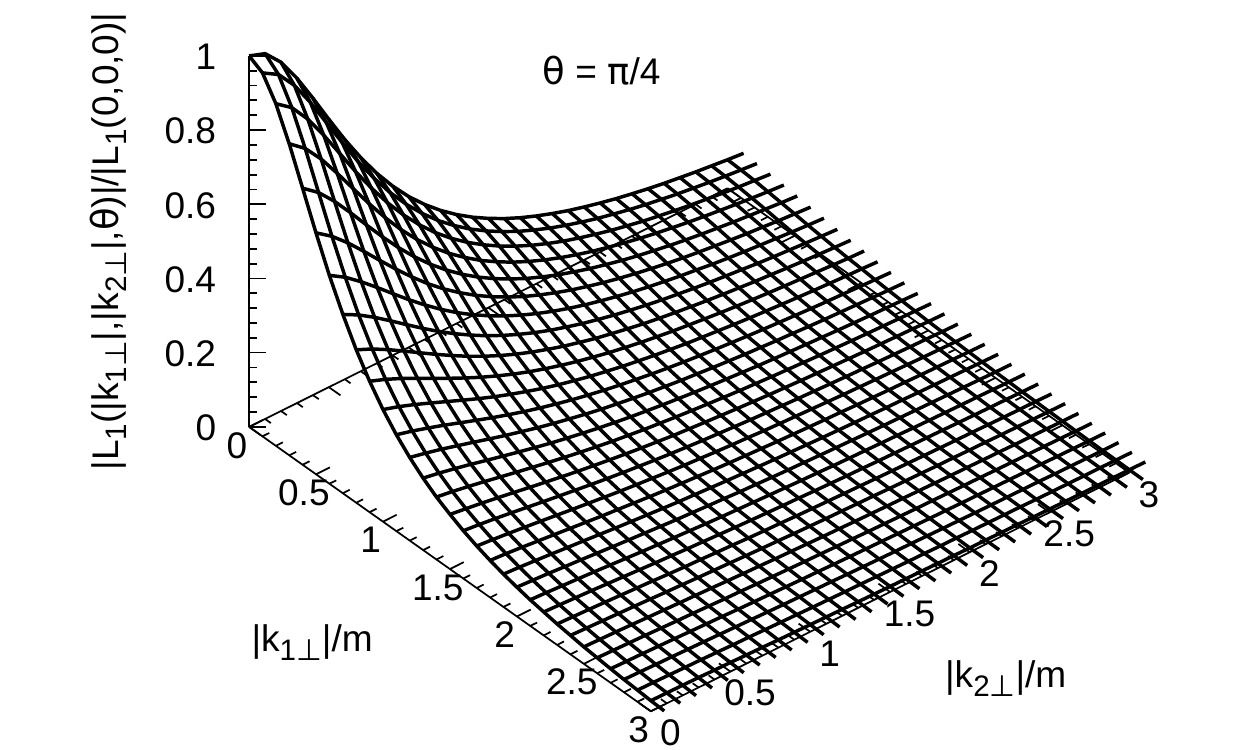}
\includegraphics[scale=0.5]{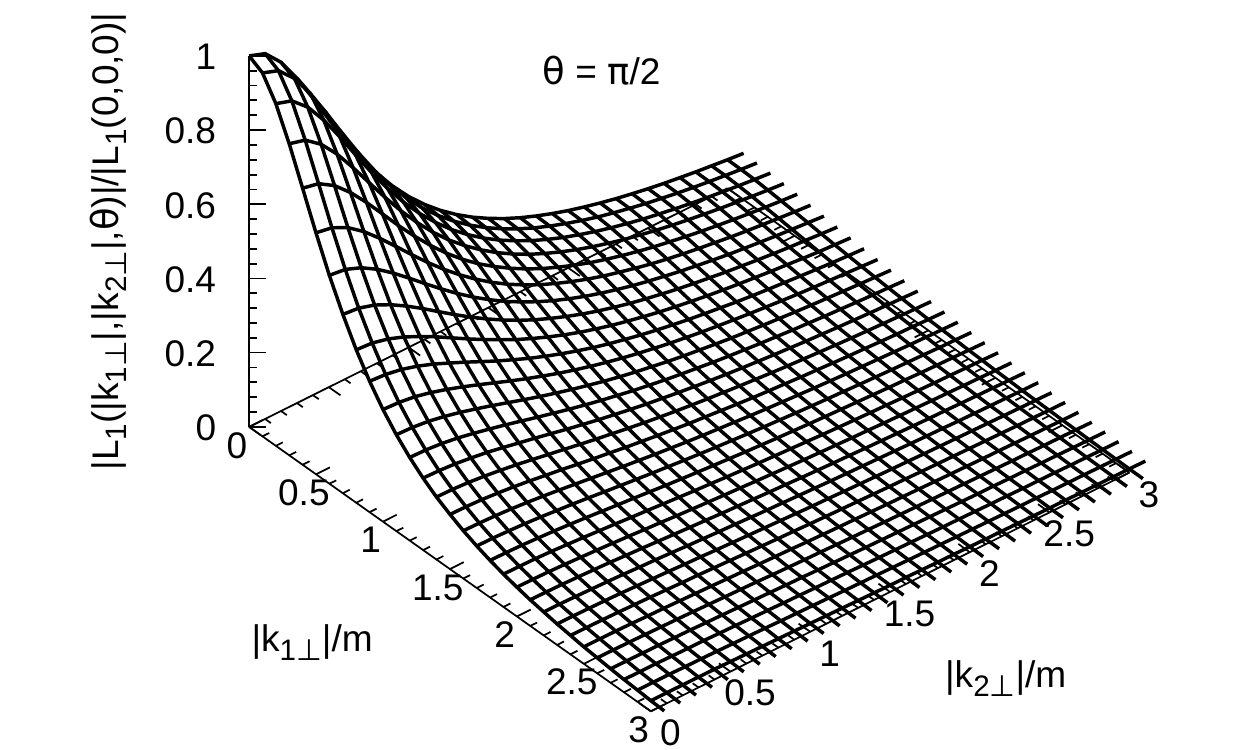}
\includegraphics[scale=0.5]{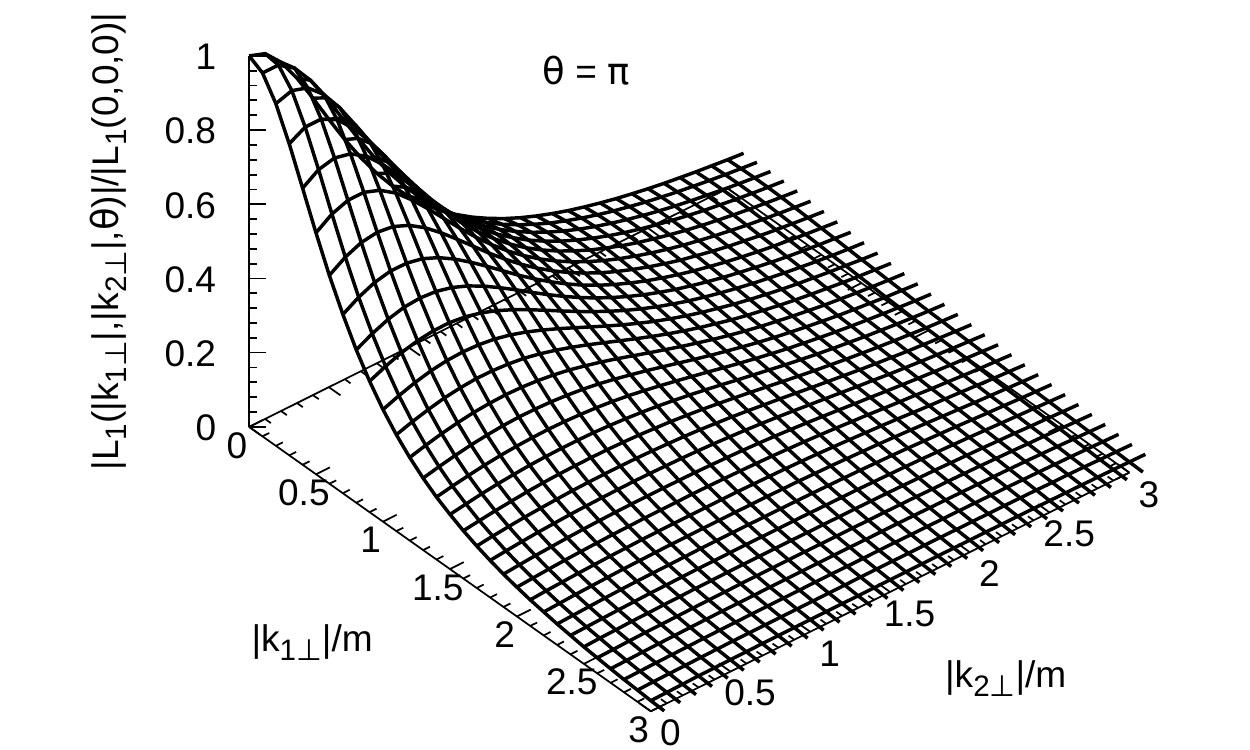}
\caption{ Transverse contribution, $L_1(|\vec{k}_{1\perp}|,|\vec{k}_{2\perp}|, \theta)$
  with respect to  $|\vec{k}_{1\perp}| $ and $|\vec{k}_{2\perp}|$, for $\theta = 0, \pi/4, \pi/2, \pi$  calculated in  Euclidean space. \label{Fig:transverse_3d}}
\end{figure}

The three-dimensional structure of the transverse amplitude is further exposed in Fig.~\ref{Fig:transverse_3d}, where we plot the absolute value of the 
 amplitude $L_1(|\vec{k}_{1\perp}|,|\vec{k}_{2\perp}|, \theta)$ with respect to $|\vec{k}_{1\perp}|$ and $|\vec{k}_{2\perp}|$ for different fixed values of $\theta$. The computations were for simplicity performed in Euclidean space, since it has already been shown above that the Euclidean and Minkowski calculations give the same results for the transverse amplitude. 
One can conclude from the figure that the transverse amplitude becomes wider when the value of is $\theta$ is increased, due to correlations proportional to $\vec{k}_{1\perp}\cdot \vec{k}_{2\perp}$. This effect is clearly visible, in particular, for the anti-aligned case $(\theta = \pi)$, i.e.~when the transverse momenta obey the relation  $|\vec{k}_{3\perp}|^2 = (|\vec{k}_{1\perp}|-|\vec{k}_{2\perp}|)^2$.   For $\theta=\pi$ and along $|\vec{k}_{1\perp}|=|\vec{k}_{2\perp}|$ there is a clear enhancement of the transverse amplitude which exhibits a bump due to the vanishing value of  $\vec k_{3\perp}$. 
The development of this pattern is seen by inspecting the evolution of the amplitude with $\theta$  and comparing the results for the $\pi/2$ and $\pi$ angles. This feature is further enhanced when the three-body bound state mass decreases for a strongly bound system. 

\subsection{Short-range correlation}

The two-body short-range correlation (in the context of non-relativistic nuclear physics, see e.g.~Ref.~\cite{HenRMP2017})  
is exhibited by  the model for large relative momentum, $|\vec{k}_{2\perp}-\vec{k}_{3\perp}|$, and a back-to-back momentum configuration of particles 2 and 3, where the transverse amplitude is dominated by the component $L_1$. Note that this amplitude is also symmetric  by the exchange of $\vec{k}_{2\perp}$ and $\vec{k}_{3\perp}$
due to the bosonic nature of the system with equal masses.
 In the relativistic context the  concept of the 
short-range two-body correlation has not yet been developed, but its  emergent imprints are observed in the transverse amplitude. 

\begin{figure}[!htbp]
\includegraphics[scale=0.6]{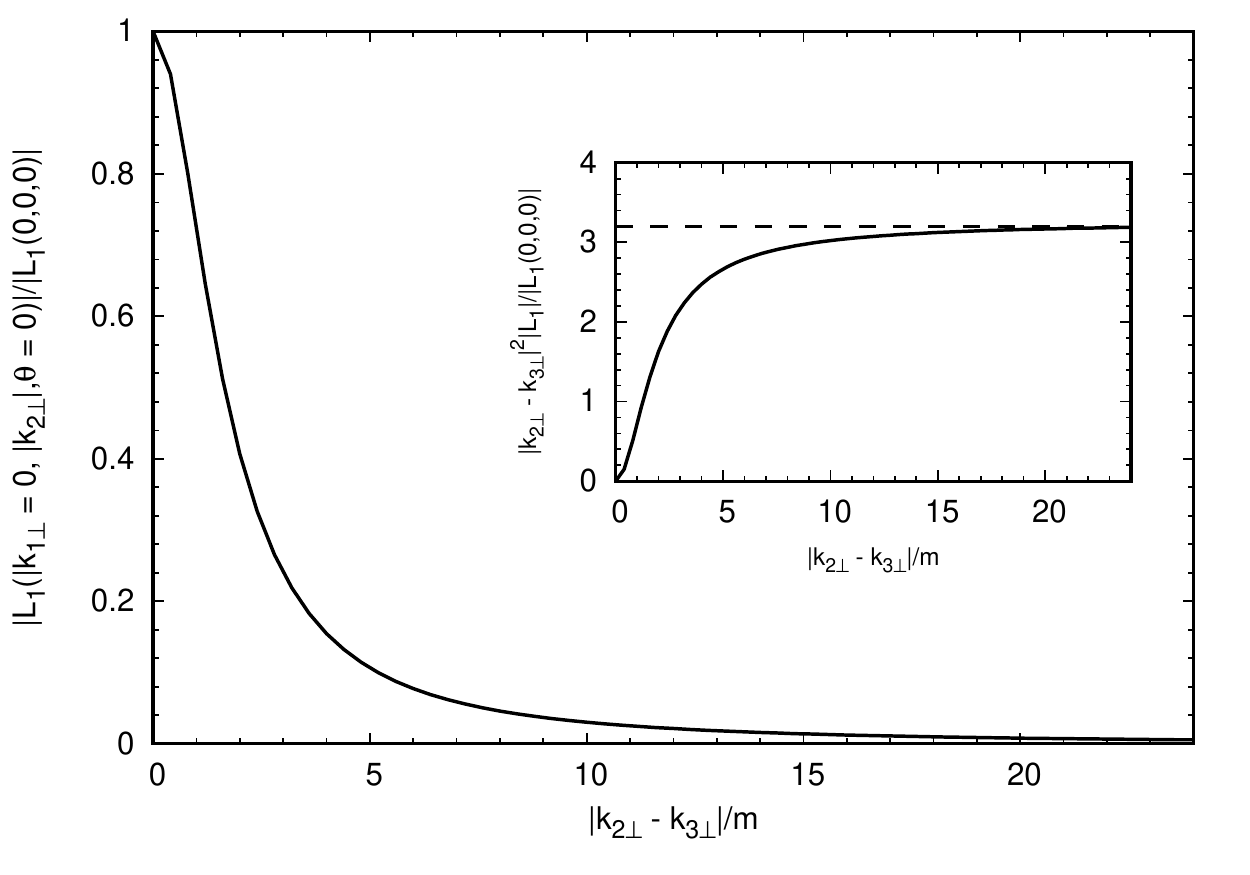}  
\caption{Transverse amplitude modulus, $|L_1|$ as a function of
   $|\vec{k}_{2\perp}-\vec{k}_{3\perp}|$  for $|\vec{k}_{1\perp}| = 0 $. In the inset  of the figure is shown $|\vec{k}_{2\perp}-\vec{k}_{3\perp}|^2|L_1|$ versus $|\vec{k}_{2\perp}-\vec{k}_{3\perp}|$ 
  with a dashed line indicating the asymptotic behavior.  \label{correl}}
\end{figure}

The plot for $L_1$ as a function of $|\vec{k}_{2\perp}-\vec{k}_{3\perp}|$ for the back-to-back configuration ($|\vec{k}_{1\perp}|=0$) is shown in Fig.~\ref{correl}. 
In the situation when $|\vec{k}_{2\perp}-\vec{k}_{3\perp}|$ is large, only $L_1$ dominates in the full three-body transverse amplitude, being the counterpart of the factorization of the non-relativistic wave function as a two-body term depending on the relative distance between the two particles times a function of  the relative coordinates of the other $N-2$ 
particles \cite{HenRMP2017}.  In the model with contact interaction, the denominator in Eq.~(\ref{Eq:chi_E_2}) provides the momentum
dependence of the ``wave function" of the short-range correlated pair in the valence LFWF of the three-body system. 
The relative momentum behavior shown in the figure reflects the free propagators of the bosons, as the contact interaction does not bring any momentum dependence. This property is shared by the non-relativistic 
three-body model with a zero-range potential. Furthermore, for $|\vec{k}_{1\perp}|=0$  the denominator of Eq.~(\ref{Eq:chi_E_2}) provides the large 
momentum behavior as $L_1\sim |\vec{k}_{3\perp}|^{-2}$ that leads to  $L_1\sim |\vec{k}_{2\perp}-\vec{k}_{3\perp}|^{-2}$ in the back-to-back configuration, as confirmed by our results. The scale for the asymptotic behavior is naturally fixed by the individual boson mass, as it can be seen in Fig.~\ref{correl}, where  $L_1$ shows an accentuated  drop in this momentum range. In the inset plot  the asymptotic behavior for large relative momentum is shown. 

The asymptotic property of $L_1$ also follows from the structure of the valence LFWF, which has the three-body LF propagator as  the dominant factor for the contact interaction and at large momentum is just the inverse of the free three-body mass, i.e.
$$M_0^2=\sum_i\frac{\vec{k}^2_{i\perp}}{\xi_i}\, .$$
 Note that
 $M_0^{-2}\sim |\vec{k}_{2\perp}-\vec{k}_{3\perp}|^{-2} $ for $|\vec{k}_{1\perp}|=0$, which is  
 the  power-law behavior seen in Fig.~\ref{correl}. Such asymptotic form is expected to change for models with finite range interactions.

\section{Conclusions}\label{Sec:Concl}

The three-body BS equation has been solved, directly in Minkowski space, by standard analytical and numerical 
methods, where ({\it i}) no ansatz or assumption has been introduced to represent 
the BS amplitude and ({\it ii}) the singularities from the kernel are treated analytically and numerically directly in the four-dimensional equation. The application of the direct integration method to the three-body Faddeev-BS equation was already presented  in Ref.~\cite{Ydrefors19_3b_Mink}. However, in the present  paper the Minkowski-space structure of the three-body system has been analyzed in far greater detail and brings up the structure of the short-range correlated pair in the Borromean system. 

The computed amplitude turns out to be highly peaked, indicating the presence of a singular behavior, as shown in Sec.~\ref{Sec:TA}. This is very different from what was found for the amplitude computed through the Wick-rotated equation in Ref.~\cite{ey3b}, due to the presence of branching points and the associated cuts. In one example we expose the cut 
contribution to the amplitude comparing results from Minkowski and Euclidean calculations.

Although the BS amplitudes obtained from the solution in Euclidean and Minkowski spaces are fundamentally different, they can be compared by means of the transverse amplitude. The comparison shows a notable agreement, giving more confidence on the reliability of the direct integration method. Furthermore,
the transverse amplitude reveals  the structure of the short-range correlated pair  in the valence wave function, which was found when the pair has large relative momentum in 
a back-to-back configuration.  We found that  the Faddeev component of the  BS 
amplitude defined with the pair interaction dominates over the others, and   a  power-law behavior of the type $\sim |\vec k_{i\perp}-\vec k_{j\perp}|^{-2}$ 
is found for the associated transverse amplitude and confirmed by our numerical results.

 In this work, we show that the results obtained by the direct integration in Minkowski space agree well with the Euclidean results for comparable quantities, however this method is quite demanding from the numerical point of view. One possible way to improve on this and additionally be able to treat more realistic kernels and/or the spin degree of freedom is to transform the BS equation into a non-singular form by using the Nakanishi integral representation~\cite{Nakanishi63,Nakanishi69}. 
 The formulation of the BS approach to the three-body problem via Nakanishi integral representation   is already in progress and computations based on this method will be undertaken in the near future. 
 Once the BS amplitude of the three-body state is known in Minkowski space it can be used 
 to investigate electromagnetic form factors, the diversity of parton longitudinal and transverse momentum distributions, as well as  the space-time 
 structure  of the pair short-range correlation.  

Furthermore, one interesting direction for future explorations of the three-body system is to consider particles with non-equal masses, as a framework, for example,
to study baryons with a heavy-light content. Of course still many steps have to be considered to include the subtle physics of Quantum-Chromodynamics (QCD) in 
a continuum model (see e.g.~Refs.~\cite{Eichmannreview,Bashir:2012fs}), but now in Minkowski space. We  expect that the formulation will also allow to explore excited states and, through it, in the low energy region, the Efimov phenomena relativistically. 

\begin{acknowledgments}
We are grateful to Jaume  Carbonell for stimulating discussions. 
We thank for the support from Conselho Nacional de Desenvolvimento Cient\'{i}fico e Tecnol\'{o}gico (CNPq) and 
Coordena\c{c}\~ao de Aperfei\c{c}oamento de Pessoal  de N\'{i}vel Superior 
(CAPES) of Brazil. J.H.A.N.~acknowledges the support of the grant  \#2014/19094-8 and V.A.K.~of the grant \#2015/22701-6 from
Funda\c{c}\~ao de Amparo \`{a} Pesquisa do Estado de S\~ao Paulo (FAPESP). E.Y.~thanks for the financial
support of the grants \#2016/25143-7 and \#2018/21758-2 from FAPESP. T.F.~thanks the FAPESP Thematic Project
grant \#17/05660-0.  V.A.K.~is also sincerely grateful to group of theoretical nuclear physics of ITA, S\~{a}o Jos\'{e} dos Campos, Brazil, for kind hospitality during his visits.
\end{acknowledgments}

\appendix
\section{Derivation of the relation between the BS amplitude and LFWF \label{App:BSA_LFWF}}

In this Appendix, we derive in detail the relation between the three-body  BS amplitude and the three-body LFWF. 

The three-body BS amplitude is defined by Eq.~\eqref{Phi0_1}. 
Let us define the integral
\begin{equation} \label{I3}
\begin{aligned}
I_3&=\int d^4x_1\, d^4x_2\, d^4x_3\,\delta (\omega \cd x_1)\ \delta (\omega\cd x_2)\ \delta (\omega\cd x_3)\
\\&\times\Phi_M(x_1,x_2,x_3;p)\exp (ik_1\cd x_1+ik_2\cd x_2+ik_3\cd x_3)\ ,
\end{aligned}
\end{equation} 
where $\omega=(\omega_0,\vec{\omega})$, $\omega^2=0$ and $k_{1,2,3}$ are the on-shell momenta, i.e.~$k_1^2=k_2^2=k_3^2=m^2$.
The delta functions in \eqref{I3} restrict the variation of the arguments of the coordinate space BS amplitude to the LF hyperplane $\omega\cd x=0$.

We now represent the $\delta $-functions in (\ref{I3}) in the integral form
\begin{equation}\label{delta}
\delta(\omega\cd x_i)=\frac{1}{2\pi}\int \exp(-i\omega\cd x_i\,\tau_i)d\tau_i. 
\end{equation}
Due to translation invariance, when all $x_i$'s are shifted by $a$: $x\to x+a$, the BS amplitude obtains a factor $\exp(-ia\cd p)$: 
\begin{eqnarray}\label{Phi1}
&&\Phi_M(x_1+a,x_2+a,x_3+a;p)= \nonumber
\\
&&=\exp(-ia\cd p)\Phi_M(x_1,x_2,x_3;p),
\end{eqnarray}
like the non-relativistic wave function.

We introduce then the BS amplitude \eqref{Phi0_1} in momentum space. We define it, extracting the delta function, responsible for conservation of momenta:
\begin{equation}\label{Phix}
\begin{aligned}
\Phi_M(&x_1,x_2,x_3;p)=\frac{(2\pi)^4}{(2\pi)^{12}}\int d^4k'_1\, d^4k'_2\, d^4k'_3\, \\
& \times\exp (-ik'_1\cd x_1-ik'_2\cd x_2-ik'_3\cd x_3)\\
&\times\delta^{(4)}(k'_1+k'_2+k'_3-p)\, \Phi_M(k'_1,k'_2,k'_3;p).
\end{aligned}
\end{equation}
Because of the delta-function in (\ref{Phix}) the amplitude $\Phi_M(x_1,x_2,x_3;p)$ satisfies the relation (\ref{Phi1}). We emphasize that here, in contrast to Eq.~(\ref{I3}), all the arguments $k'_{1,2,3}$  of the BS amplitude are the off-mass-shell momenta.

We substitute (\ref{Phix}) and (\ref{delta}) in (\ref{I3}) and integrate over $x_{1,2,3}$ and $k'_{1,2,3}$. The result reads:
\begin{equation}\label{I3a}
\begin{aligned}
I_3&=\frac{1}{(2\pi)^{3}}\int d\tau_1 d\tau_2 d\tau_3\,(2\pi)^4 \\
&\times\delta(k_1+k_2+k_3-p-\omega\tau_1-\omega\tau_2-\omega\tau_3)\\
&\times\Phi_M(k_1-\omega\tau_1,k_2-\omega\tau_2,k_3-\omega\tau_3;p) \, ,
\end{aligned}
\end{equation}
where the variables  $\tau_{1,2,3} \in\, ]-\infty,\infty [ $.
To avoid confusion, we emphasize again: the four-momenta $k_{1,2,3}$ in this formula are the on-shell momenta, according to the definition of (\ref{I3}),
whereas, the arguments of the BS amplitude are the off-shell momenta, like $k'_{1,2,3}$ in (\ref{Phix}); namely: $(k_1-\omega\tau_1)^2=m^2-2(\omega\cd k_1)\tau_1\neq m^2$ since $\tau_1\neq 0$ and similarly for other arguments except for $p^2=M_3^2$.


We subsequently introduce the variable $\tau=\tau_1+\tau_2+\tau_3$ 
and represent the integral (\ref{I3a}) in the following three equivalent forms:
\begin{small}
\begin{equation}\label{I3b}
\begin{aligned}
&I_{3a}=\frac{(2\pi)^4}{(2\pi)^{3}}\int  d\tau\, \delta(k_1+k_2+k_3-p-\omega\tau)\\
& \times\int d\tau_1 d\tau_2 \\
&\times\Phi_M(k_1-\omega\tau_1,k_2-\omega\tau_2,k_3-\omega(\tau-\tau_1-\tau_2);p),\\
&I_{3b}=\frac{(2\pi)^4}{(2\pi)^{3}}\int d\tau\, \delta(k_1+k_2+k_3-p-\omega\tau)\\
&\times\int d\tau_2 d\tau_3 \\
&\times\Phi_M(k_1-\omega(\tau-\tau_2-\tau_3),k_2-\omega\tau_2,k_3-\omega\tau_3;p),\\
&I_{3c}=\frac{(2\pi)^4}{(2\pi)^{3}}\int d\tau\,\delta(k_1+k_2+k_3-p-\omega\tau)\\
&\times\int d\tau_1 d\tau_3 \\
&\times\Phi_M(k_1-\omega\tau_1,k_2-\omega(\tau-\tau_1-\tau_3),k_3-\omega\tau_3;p).
\end{aligned}
\end{equation}
\end{small}
It can be also represented as:
\begin{equation}\label{I3c}
\begin{aligned}
I_3&=\frac{(2\pi)^4}{(2\pi)^{3}}\int d\tau\, \delta(k_1+k_2+k_3-p-\omega\tau)\\
&\times\int d\tau_1 d\tau_2 d\tau_3\;\delta(\tau_1+\tau_2+\tau_3-\tau)\\
&\times \Phi_M(k_1-\omega\tau_1,k_2-\omega\tau_2,k_3-\omega\tau_3;p),
\end{aligned} 
 \end{equation}
where by means of the delta function $ \delta(\tau_1+\tau_2+\tau_3-\tau)$ one can exclude any $\tau_i$ and obtain Eq.~(\ref{I3b}).
Depending on the convenience, one can chose any of the forms in (\ref{I3b}) to calculate the double integral over $\tau_i,\,\tau_j$. With the standard choice
$\omega^\mu=(1,0,0,-1)$, i.e., in the LF coordinates, $\vec{\omega}_{\perp}=0,\; \omega^+=0,\; \omega^-=2$, these integrals are reduced to the integrals over the $k^-_{i}$ components. The value of $\tau$ is determined from the conservation law $k_1+k_2+k_3=p+\omega\tau$. For example, squaring this equation, we find
\begin{equation}
\begin{aligned}
\tau&=\frac{(k_1+k_2+k_3)^2-M_3^2}{2(\omega\cd p)}=\\
&\frac{1}{2(\omega\cd p)}
\left(\frac{k_{1\perp}^2+m^2}{x_1}
+\frac{k_{2\perp}^2+m^2}{x_2}+\frac{k_{3\perp}^2+m^2}{x_3}-M_3^2\right)\\
&\equiv\frac{1}{2(\omega\cd p)}\left(M_0^2-M_3^2\right).
\end{aligned}
\end{equation}

On the other hand, the integral (\ref{I3}) can be expressed in terms  of the
three-body LFWF. We assume that the  LF plane is
the limit of a space-like plane, therefore  the  operators $\varphi (x_1),\varphi (x_2),\varphi (x_3),$ commute with each other, and, hence,  the  symbol of the $T$ product in
(\ref{Phi0_1}) can be omitted. In the  considered representation, the Heisenberg
operators $\varphi (x)$ in  (\ref{Phi0_1}) are identical on the light front
$\omega\cd x=0$ to the  Schr\"odinger ones (just as in the ordinary
formulation of field  theory the Heisenberg and Schr\"odinger operators are
identical for  $t=0$).  The Schr\"odinger operator $\varphi (x)$ (for the
spinless  case, for simplicity), which for $\omega\cd x=0$ is the free field
operator, is given by:
\begin{equation} \label{bs5}
\begin{aligned}
\varphi(x)&=\frac 1{(2\pi )^{3/2}}\int \frac{d^3k}{\sqrt{2\varepsilon_k}}\\
&\times\left[a(\vec k)
\exp (-ik\cd  x)+a^{\dagger}(\vec k)\exp (ik\cd  x)\right].
\end{aligned}
\end{equation}
We represent the state vector $|p\rangle \equiv \phi (p)$ in  (\ref{I3}) in
the form of the expansion via the Fock states:
\begin{equation}\label{wfp1}
\begin{aligned}
|p\rangle&=(2\pi)^{3/2}\int\psi(k_1,k_2,k_3,p,\omega\tau)\\
&\times\delta^{(4)}(k_1+k_2+k_2-p-\omega\tau)2(\omega\cd p)d\tau \\
&\times
 \frac{d^3k_1}{(2\pi)^{3/2}\sqrt{2\varepsilon_{k_1}}}
\frac{d^3k_2}{(2\pi)^{3/2}\sqrt{2\varepsilon_{k_2}}}
\frac{d^3k_3}{(2\pi)^{3/2}\sqrt{2\varepsilon_{k_3}}}\\
&\times a^\dagger(\vec{k}_1)a^\dagger(\vec{k}_2)
a^\dagger(\vec{k}_3)|0\rangle  + \cdots\ .
\end{aligned}
\end{equation}
In (\ref{bs5}) and (\ref{wfp1}) the four-momenta $k_{1,2,3}$ are on mass-shells. We substitute this expression in \\
$\Phi_M(x_1,x_2,x_3;p)$, Eq.~(\ref{Phi0_1}). Since the vacuum  state on the light
front is always ``bare'', the creation operator,  applied to the vacuum state
$\langle 0|$ gives zero, and in the  operators $\varphi (x)$  the part
containing the annihilation  operators only survives. This cuts out the
three-body Fock component in  the  state vector.  We thus obtain:
\begin{equation}
\begin{aligned}
\Phi_M(&x_1,x_2,x_3;p)=(2\pi)^{3/2}\int\psi(k_1,k_2,k_3,p,\omega\tau)
\\
&\times\delta^{(4)}(k'_1+k'_2+k'_2-p-\omega\tau)2(\omega\cd p)d\tau\\
&\times\exp(-ik'_1x_1-ik'_2x_2-ik'_3x_3)\\
&\times\frac{d^3k'_1}{(2\pi)^{3}2\varepsilon_{k'_1}}\frac{d^3k'_2}{(2\pi)^{3}2\varepsilon_{k'_2}}\frac{d^3k'_3}{(2\pi)^{3}2\varepsilon_{k'_3}}.
\end{aligned}
\end{equation}
Then we substitute this $\Phi_M(x_1,x_2,x_3;p)$
in (\ref{I3}) and integrate over $x_{1,2,3}$. The integration, for example, over $x_1$ and then over $k'_1$  is fulfilled as follows 
\begin{equation}
\begin{aligned}
&\int \frac{d^4x_1 d^3k'_1}{(2\pi)^{4}2\varepsilon_{k'_1}} \exp(-ik'_1x_1)\exp(-i\omega\cd x_1\tau_1)d\tau_1 \\
&\times\exp(ik_1x_1)\\
&=\int \delta^{(4)}(k_1-k'_1-\omega\tau_1)d\tau_1\frac{d^3k'_1}{2\varepsilon_{k'_1}}
\\
&=\int \delta^{(4)}(k_1-k'_1-\omega\tau_1)d\tau_1d^4k'_1\theta(\omega\cd k'_1)\delta({k'}_1^2-m^2)\\
&=\int \theta(\omega\cd k_1)\delta((k_1+\omega\tau_1)^2-m^2)\\
&=\int d\tau_1 \delta(2(\omega \cd k_1)\tau_1)
=\frac{1}{2(\omega \cd k_1)},
\end{aligned}
\end{equation}
and similarly for the integrations over $x_2,k'_2$ and $x_3,k'_3$.  We used here that $\theta(k'_{10})=\theta(\omega\cd k'_1)$ for  $k'_{10}>0$.
Then for $I_3$ we get (cf.~Eq.~(3.56) from Ref.~\cite{cdkm}):
\begin{equation} \label{bs6}
\begin{aligned}
I_3&=\frac{(2\pi )^{3/2}2(\omega\cd p)}{2(\omega\cd k_1)2(\omega\cd k_2)2(\omega\cd k_3)}
\int d\tau\psi (k_1,k_2,k_3;p,\omega \tau )\\
&\times\delta^{(4)}(k_1+k_2+k_3-p-\omega \tau )\ .
\end{aligned}
\end{equation}
Comparing (\ref{bs6}) and (\ref{I3c}), we find:
\begin{equation} \label{bs7}
\begin{aligned}
\psi (k_1&,k_2,k_3,p,\omega \tau )=\frac{1}{\sqrt{2\pi}}\frac{2(\omega\cd k_1)2(\omega\cd k_2)2(\omega\cd k_3)}{2(\omega \cd p)}\\
&\times\int d\tau_1 d\tau_2 d\tau_3\;\delta(\tau_1+\tau_2+\tau_3-\tau)\\
&\times \Phi_M(k_1-\omega\tau_1,k_2-\omega\tau_2,k_3-\omega\tau_3;p).
\end{aligned} 
\end{equation}
As mentioned, in ordinary LFD, Eq.~(\ref{bs7}) corresponds to the integration
over $k^{-}$. This equation makes the  link between the three-body BS amplitude $\Phi_M$ and the wave function  $\psi$ defined on the light front specified by $\omega$. As it is seen from the above derivation, it is generalizable (with the same coefficient $1/\sqrt{2\pi})$ for arbitrary number of particles. Simply the number of the factors $2(\omega\cd k_i)$ and of the arguments increases. 
In the LF coordinates Eq.~(\ref{bs7}) obtains the form: 
\begin{small}
\begin{multline} \label{bs7a}
\psi (\vec{k}_{1\perp},\xi_1;\vec{k}_{2\perp},\xi_2;\vec{k}_{3\perp},\xi_3)=
 \frac{1}{\sqrt{2\pi}}\frac{(p^+)^2}{2} 2\xi_1 2\xi_2 2\xi_3 \\ \times \int d\tau_1 d\tau_2 d\tau_3\delta(\tau_1+\tau_2+\tau_3-\tau)
\Phi_M(\tilde k_1;\tilde k_2;\tilde k_3;p).
\end{multline}
\end{small}
where $\tilde k_i\equiv\{ \vec{k}_{i\perp},k^+_{i},k^-_{i}-2\tau_i\}$ and $0<\xi_i<1$ $(\xi_1+\xi_2+\xi_3=1)$ denotes the longitudinal momentum fraction of particle $i$.

We introduce now new integration variables: \\ $k'^-_{1}=k^-_{1}-2\tau_1$, etc, and then
\begin{equation} \label{bs7b}
\begin{aligned}
&\psi (\vec{k}_{1\perp},\xi_1;\vec{k}_{2\perp},\xi_2;\vec{k}_{3\perp},\xi_3)=\frac{(p^+)^2}{\sqrt{2\pi}} \, \xi_1 \xi_2 \xi_3\\
&\times\int dk^{-}_{1}\, dk^{-}_{2}\, 
\Phi_M(k_1,k_2,k_3;p)\\
&=\frac{(p^+)^2}{\sqrt{2\pi}}  \xi_1 \xi_2 \xi_3
\int dk^-_{1}dk^-_{2} 
\Phi_M(k_1,k_2,k_3;p).
\end{aligned}
\end{equation}
In the last line, we  omitted the integration over the 3rd argument since it is not independent. In the above formula it is understood that $k_3=p-k_1-k_2$. 
One can chose any pair of arguments: (12), (13) or (23), depending on the convenience.

\section{Calculating the Euclidean transverse amplitude}\label{App:transv_eucl_kernel}

In this Appendix we derive in detail the function $\chi$ occurring in the expression for the Euclidean transverse amplitude, Eq.~\eqref{Eq:L1_E_final}.

From Eqs.~\eqref{Eq:phi_E} and \eqref{Eq:L_E}, we define $\chi$ as the integral
\begin{equation}\label{Eq:chi_E}
\begin{aligned}
\chi(k'_{14},&k'_{1z}; \vec{k}'_{1\perp}, \vec{k}'_{2\perp}) =  \\
&\int_{-\infty}^{\infty} dk'_{20} \int_{-\infty}^{\infty} dk'_{2z} \frac{i}{(k'_{24}-i\frac{M_3}{3})^2+k'^2_{2z}+m^2_2}\\
&\times\frac{i}{(k'_{14}+k'_{24}+i\frac{M_3}{3})^2+(k'_{1z}+k'_{2z})^2+m^2_3}. 
\end{aligned}
\end{equation}

The two propagators in \eqref{Eq:chi_E} can then  be put together by using the Feynman parametrization \eqref{Eq:Feynman_param}
leading to the result
\begin{equation}
\label{Eq:denom_prod}
\begin{aligned}
&\frac{i}{(k'_{24}-i\frac{M_3}{3})^2+k'^2_{2z}+m^2_2}\\
&\times\frac{i}{(k'_{14}+k'_{24}+i\frac{M_3}{3})^2+(k'_{1z}+k'_{2z})^2+m^2_3}\\
&= - \int_{0}^1 \frac{du}{D^2},
\end{aligned}
\end{equation}
where the denominator reads
\begin{equation}
\label{Eq:D_eucl}
\begin{aligned}
D &= k'^2_{24}+k'^2_{2z} + (1-u)\bigl[k'^2_{14} + k'^2_{1z}\bigr]  + \frac{2}{3}iM_3k'_{24} \\
& + 2(1-u)k'_{1z}k'_{2z} + \frac{2}{3}(1-u)k'_{14}(3k'_{24}+iM_3) \\
& - \frac{4}{3}iuM_3k'_{24} + (1-u)m^2_3 + u\,m^2_2  - \frac{M^2_3}{9} = \\
& k'^2_{24}+k'^2_{2z} + (1-u)\bigl[k'^2_{14} + k'^2_{1z}\bigr] + \\
& 2\Bigl[(1-u)k'_{14}-\frac{iM_{3}}{3}(-1+2u) \Bigr]k'_{24} \\
& + 2(1-u)k'_{1z}k'_{2z} + \frac{2}{3}iM_3(1-u)k_{14} + (1-u)m^2_3 \\
&+u\,m^2_2 - \frac{M^2_3}{9}.
\end{aligned}
\end{equation}

We subsequently eliminate the terms linear in $k'_{24}$ and $k'_{2z}$, by performing in Eqs.~\eqref{Eq:chi_E}, \eqref{Eq:denom_prod} and \eqref{Eq:D_eucl}  the transformations
\begin{equation}
\begin{aligned}
k'_{24} &\longrightarrow k'_{24} - \alpha,\\
k'_{2z} &\longrightarrow k'_{2z} -\beta,
\end{aligned}
\end{equation}
with 
\begin{equation}
\alpha = (1-u)k'_{14}-\frac{iM_3}{3}(-1+2u),  
\end{equation}
and
\begin{equation}
\beta=(1-u)k'_{1z}.
\end{equation}

By these transformations the denominator \eqref{Eq:D_eucl} is changed into
\begin{equation}
\begin{aligned}
D\longrightarrow\tilde{D} = k'^2_{24}+k'^2_{2z} + A,  
\end{aligned}
\end{equation}
where
\begin{equation}
\begin{aligned}
A &=  u(1-u)\bigl[k'^2_{14} + k'^2_{1z}\bigr] + (1-u)m^2_3 + um^2_2 \\
& + \frac{4}{3}iM_3u(1-u)k'_{14} - \frac{4}{9}M^2_3u(1-u).  
\end{aligned}
\end{equation}

The integrals over $k'_{24}$ and $k'_{2z}$ in \eqref{Eq:chi_E} can now be performed analytically, and the result is
\begin{equation}
\label{Eq:chi_E_2_new}
\begin{aligned}
\chi(&k'_{14},k'_{1z}; \vec{k}'_{1\perp}, \vec{k}'_{2\perp}) = -\int_{0}^1 du\\
&\times\int_{-\infty}^{\infty} dk'_{20} \int_{-\infty}^{\infty} \frac{dk'_{2z}}{(k'^2_{24}+k'^2_{2v}+A)^2}= \\
& -2\pi \int_{0}^1 du \int_{0}^{\infty} \frac{k'dk'}{(k'^2+A)^2}=-\pi\int_{0}^1 \frac{du}{A}. 
\end{aligned}
\end{equation}

Alternatively, one can write the quantity $A$ in the form
\begin{equation}
A = au^2+bu+c,
\end{equation}
with
\begin{equation}
\label{abc_2}
\begin{aligned}
a&=-k'^2_{1z}-\Bigl(k'_{14}+\frac{2}{3}iM_3\Bigr)^2,\\
b&=k'^2_{1z}+\Bigl(k'_{14}+\frac{2}{3}iM_3\Bigr)^2+m^2_2-m^2_3,\\
c&=m^2_3.
\end{aligned}
\end{equation}

\section{Numerical methods}\label{Sec:NM}

We solve in this work Eq.~\eqref{Eq:v} by expanding the amplitude $v_M(q_0,q_v)$ in a bicubic spline basis, on a finite domain $\Omega=I_{q_0}\times I_{q_v}=[-q^{\text{max}}_{0},q^{\text{max}}_{0}]\times [0,q^{\text{max}}_v]$, i.e.~
\begin{equation}
v_M(q_0,q_v)=\sum_{k=0}^{2N_{q_0}+1}\sum_{l=0}^{2N_{q_v}+1}A_{ij}S_k(q_0)S_l(q_v),
\label{Eq:v_exp}
\end{equation}
where the unknown coefficients $A_{ij}$ are to determined. In the numerical implementation, the interval $I_x\:(x=q_0,q_v)$ is partioned into $N_x$ subintervals, so that good convergence was reached. The adopted spline functions, $S_j(x)$ are given by \cite{ckfb}
\begin{equation}
\begin{aligned}
S_{2i}(x)&=\left\lbrace \begin{array}{l l}
3\left(\frac{x-x_{i-1}}{h_i}\right)^2-2\Bigl(\frac{x-x_{i-1}}{h_i}\Bigr)^3,  
\\[1.2\medskipamount] \quad \text{if } x\in[x_{i-1},x_i]\\[1.7\medskipamount]
3\left(\frac{x_{i+1}-x}{h_{i+1}}\right)^2-2\left(\frac{x_{i+1}-x}{h_{i+1}}\right)^3,  \\[1.2\medskipamount]
\quad \text{if } x\in[x_i,x_{i+1}]\\[1.7\medskipamount]
0, \quad \text{if } x\not\in[x_{i-1},x_{i+1}]
\end{array}\right.\\[1.2\medskipamount]
S_{2i+1}(x)&=\left\lbrace \begin{array}{l l}
\left[-\left(\frac{x-x_{i-1}}{h_i}\right)^2+\left(\frac{x-x_{i-1}}{h_i}\right)^3\right]h_i,  \\[1.2\medskipamount]
\quad \text{if } x\in[x_{i-1},x_i]\\[1.7\medskipamount]
\left[\left(\frac{x_{i+1}-x}{h_{i+1}}\right)^2-\left(\frac{x_{i+1}-x}{h_{i+1}}\right)^3\right]h_{i+1}, \\[1.2\medskipamount]
\quad  \text{if } x\in[x_i,x_{i+1}]\\[1.7\medskipamount]
0, \quad  \text{ if } x\not\in[x_{i-1},x_{i+1}]
\end{array}\right.\\
\end{aligned}
\end{equation} 
with $h_{i}=x_{i}-x_{i-1}$.

By using \eqref{Eq:v_exp}, Eq.~\eqref{Eq:v} can be transformed to a generalized eigenvalue problem of the form
\begin{equation}
\label{Eq:EVEQ}
\sum_{i'j'}F_{iji'j'}A_{i'j'}=\lambda(M_3)\sum_{i'j'}V_{iji'j'}A_{i'j'},
\end{equation}
where
\begin{equation}
F_{iji'j'}=S_{i'}(q^{(i)}_{0})S_{j'}(q^{(j)}_{v}),
\end{equation}
and the array $V_{iji'j'}$ is the right-hand side of \eqref{Eq:v} with $v_M$ replaced by $S_{i'}(q^{(i)}_{0})S_{j'}(q^{(j)}_{v})$. The variable $q_0$ ($q_v$) has here been discretized on a mesh consisting of $2N_{q_0}+2$ ($2N_{q_v}+2$) points. The three-body mass $M_3$, or equivalently the three-body binding energy $B_3$, can subsequently be obtained from the condition
\begin{equation}
\label{Eq:Lamba_M3}
\lambda(M_3)=1.
\end{equation} 
Eq.~\eqref{Eq:Lamba_M3} constitutes a non-linear equation relative to $M_3$ and is rather time-consuming to solve. For simplicity, we use thus instead as inputs in the calculations the scattering length $a$ and the $M_3$, obtained from the solution of the Euclidean BS equation. Eq.~\eqref{Eq:EVEQ} is then solved for the eigenvalue $\lambda$ and the coefficients $A_{ij}$.

The kernel $\Pi(q_0,q_v,k_0,k_v)$ (see Eq.~\eqref{Eq:PI}), which enters Eq.~\eqref{Eq:v} has logarithmic singularities, and the analytic expressions for the singular points are given by Eqs.~\eqref{Eq:sing_k0} and \eqref{Eq:sing_k}. In the present work, the integrals over $k_0$ and $k_v$ are computed  by dividing a given integration interval into subintervals $I_i=[a_i,b_i]$, so that each subinterval contains at most one singular point which is just one of the end points of the subinterval. For each subinterval, the integrand
singularity is subsequently weakened by adopting a change of variables of the form 

\begin{equation}
\int_{a_i}^{b_i}f(x)dx=\int_0^{\sqrt{b_i-a_i}}2tf(a_i+t^2)dt, 
\end{equation}   
 for a subinterval with a singularity at $a_i$, and
\begin{equation}
\int_{a_i}^{b_i}f(x)dx=\int_0^{\sqrt{b_i-a_i}}2tf(b_i-t^2)dt, 
\end{equation}   
if the singularity is at the end point $b_i$. The resulting integrals involving smooth functions can then be performed by Gauss-Legendre integration.

\begin{figure}[bth]
\centering
\includegraphics[scale=0.25]{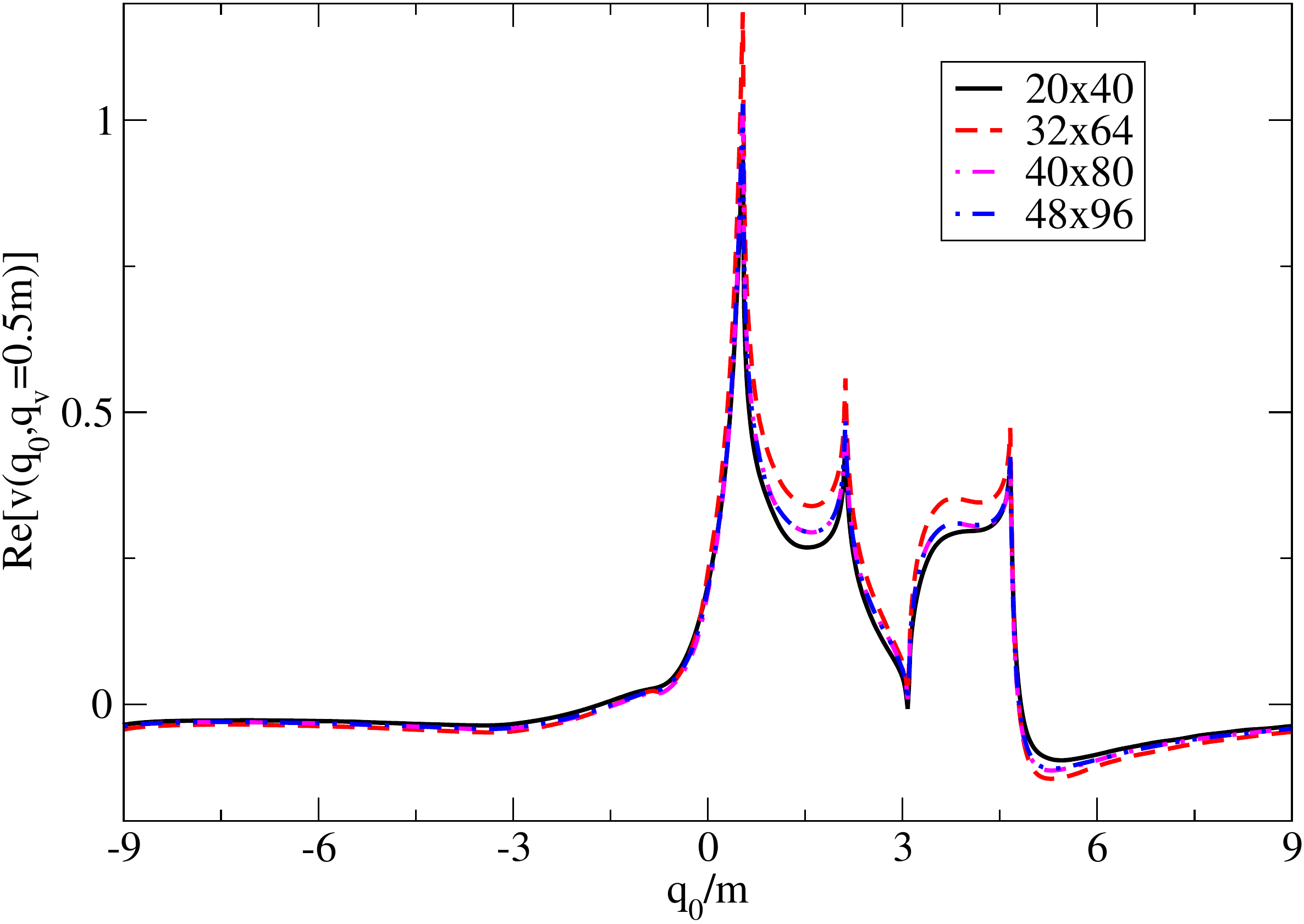}  \includegraphics[scale=0.25]{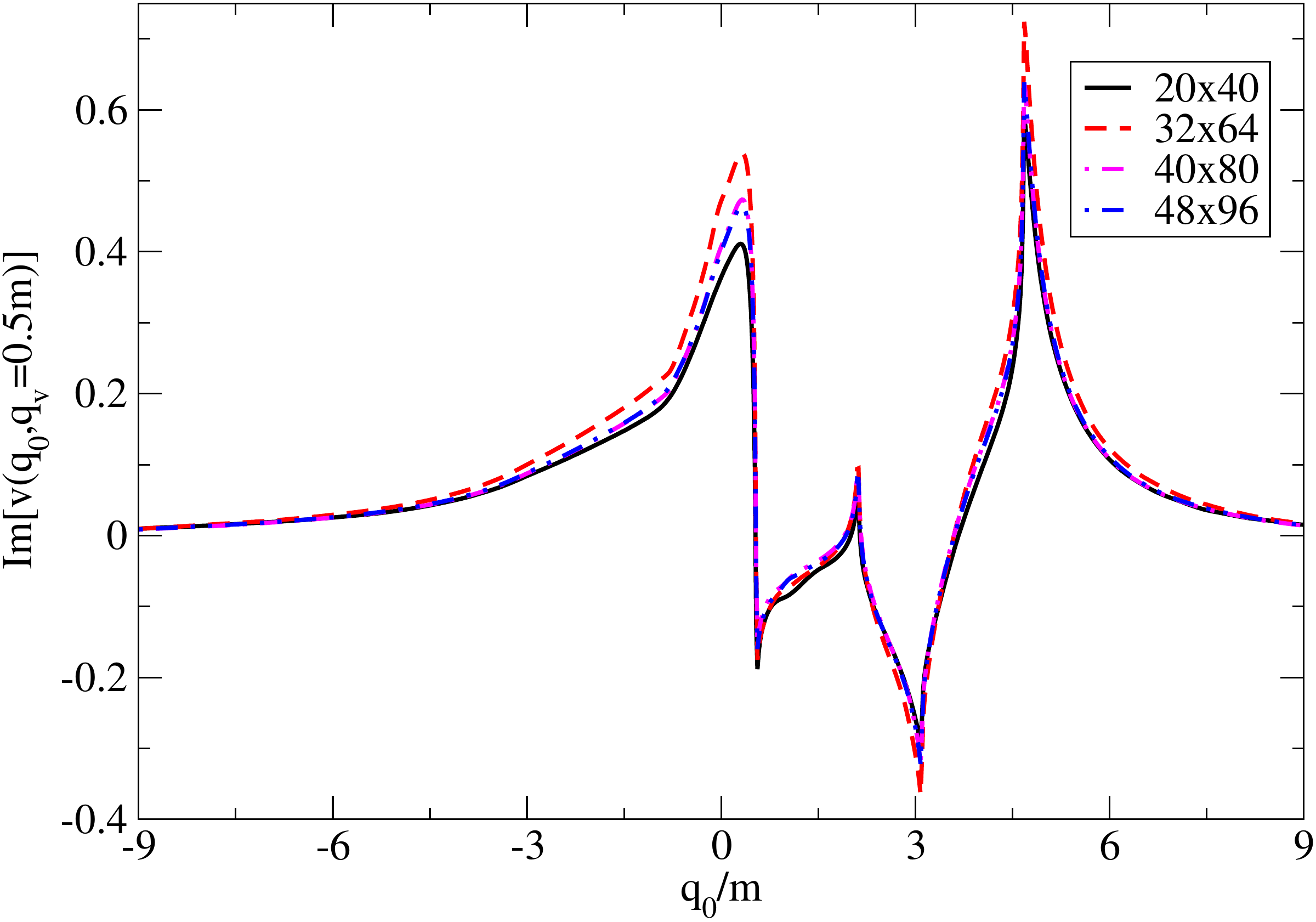} 
\caption{Convergence of the real (left panel) and imaginary (right panel) parts of the vertex function $v_M(q_0,q_v=0.5m)$ with respect to the size of the basis, $N_{q_v}\times N_{q_0}$. In the calculations we used $B_3/m=0.395$.\label{Fig:Convergence}}
\end{figure} 

\subsection{Numerical convergence}
As mentioned, in this work the three-body BS equation is solved by using an expansion of the amplitude $v_M(q_0,q_v)$ in terms of a finite number of spline functions. Evidently, it is important to check that the adopted number basis functions is enough.

For this purpose, we show in Fig.~\ref{Fig:Convergence} the real and imaginary parts of $v_M(q_0,q_v=0.5m)$, computed by using different number of subintervals $N_{q_v}$ and $N_{q_0}$, corresponding to the variables $q_v$ and $q_0$. In the calculations we used the parameters $am=-1.5$ and $B_3/m=0.395$. It is seen in the figure that for $N_{q_v}\geq 40$ and $N_{q_0}\geq 80$, the solution is well-converged.

\begin{center}
\begin{figure}[!htbp]
\centering
\includegraphics[scale=0.3]{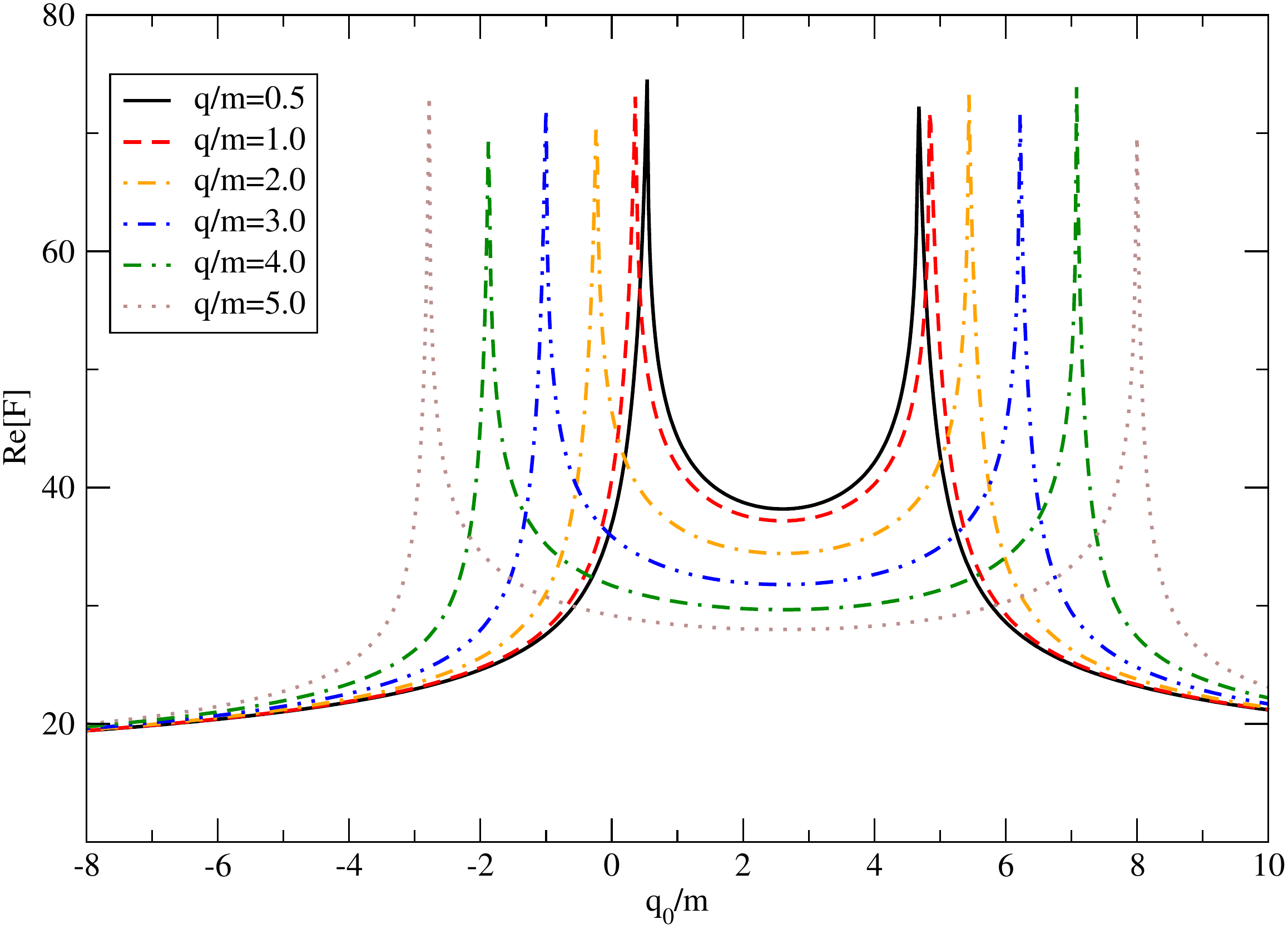} 
\includegraphics[scale=0.3]{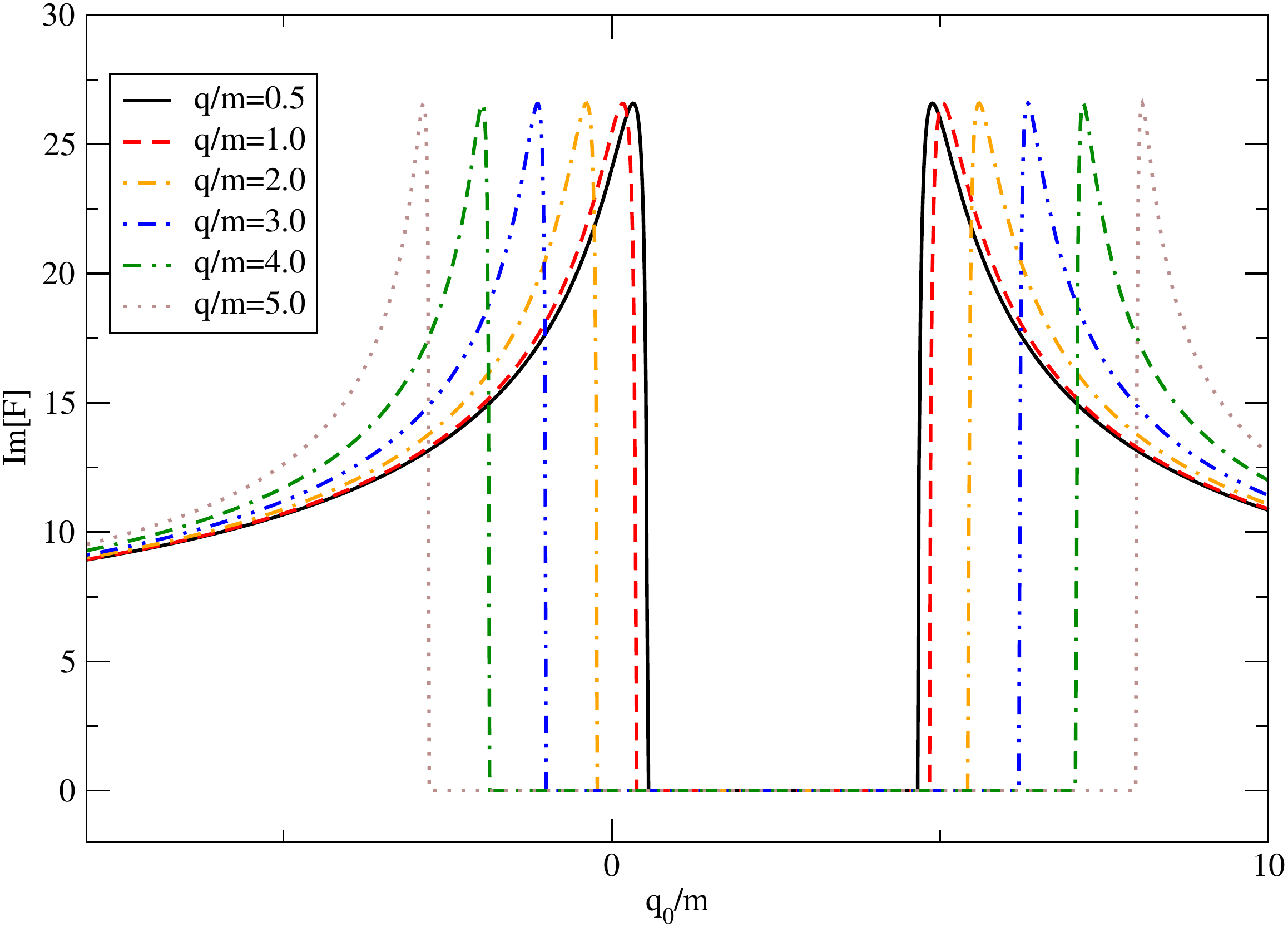}
\caption{Real and imaginary parts of ${\cal F}(M^2_{12})$ with respect to $q_0$ for different fixed values of $q_v$. \label{Fig: F_amp}}
\end{figure}
\end{center}

\subsection{Behavior of ${\cal F}(M^2_{12})$}\label{Sec:F_num}

For negative $a$, the function ${\cal F}(M^2_{12})$ is non-singular and continuous. However, the function may change rapidly in the neighbourhood of the transition points $M_{12}^2=0$ and $M_{12}^2=4m^2$. In terms of $q_0$ (for a given $q_v$) these are
\begin{equation}
q_0=M_3\pm q_v, \quad q_0=M_3\pm \sqrt{q_v^2+4m^2}. 
\end{equation}

In the Fig.~\ref{Fig: F_amp} the real and imaginary parts of $\cal F$ are shown as functions of $q_0$ (for selected values of $q_v$) in the case of $M_3/m=2.605$ corresponding to $am=-1.5$. It is seen in the figures that close to $q_0=M_3\pm \sqrt{q_v^2+4m^2}$ (i.e. $M_{12}^2=4m^2$), the amplitude has a non-smooth behavior. 

Although the non-smoothness exists, this was \\ shown to not be problematic in solving the equation. To show that, we tested solving the problem proposing a factorization of the form
\begin{equation}
v_M(q_0, q_v)={\cal F}(M^2_{12}(q_0,q_v)) \, \tilde{v}_M(q_0,q_v),
\end{equation}
by introducing
\begin{equation}
\tilde{\Pi}(q_0,q_v;k_0,k_v) = {\cal F}(M^2_{12}(q_0,q_v)) \, \Pi(q_0,q_v;k_0,k_v), 
\end{equation}
and obtaining an integral equation in terms of the function $\tilde{v}_M$ instead of $v_M$. The resulting equation was solved by expanding $\tilde{v}_M$ in splines. The result showed no significant difference between the solutions with and without the decomposition, with the convergence being achieved with a similar set of basis functions and integration points.

\end{document}